\RequirePackage{lineno}
\documentclass[twocolumn,preprintnumbers,footinbib,amsmath,amssymb,aps,groupedaddress,noeprint]{revtex4}

\usepackage[colorlinks,citecolor=blue,linkcolor=red,hyperindex,CJKbookmarks,driverfallback=dvipdfm]{hyperref}
\usepackage{amsmath}
\usepackage{graphicx}
\usepackage{dcolumn}
\usepackage{siunitx}
\usepackage{mathrsfs}
\usepackage{amssymb}
\usepackage{amsfonts}

\usepackage{natbib}
\usepackage{hyperref}
\usepackage[normalem]{ulem}

\usepackage{lineno}
\let\oldequation\equation
\let\oldendequation\endequation

\renewenvironment{equation}
{\linenomathNonumbers\oldequation}
{\oldendequation\endlinenomath}

\begin{document}

\title{Entanglement-interference complementarity and experimental
	demonstration in a superconducting circuit}
\author{Xin-Jie Huang$^{1}$}
\thanks{These authors contributed equally to this work.}
\author{Pei-Rong Han$^{1}$}
\thanks{These authors contributed equally to this work.}
\author{Wen Ning$^{1}$}
\author{Shou-Bang Yang$^{1}$}
\author{Xin Zhu$^{1}$}
\author{Jia-Hao L$\ddot{u}$$^{1}$}
\author{Ri-Hua Zheng$^{1}$}
\author{Hekang Li$^{2}$}
\author{Zhen-Biao Yang$^{1}$}
\email[Corresponding author, e-mail: ]{zbyang@fzu.edu.cn}
\author{Kai Xu$^{2,3}$}
\author{Chui-Ping Yang$^{4}$}
\email[Corresponding author, e-mail: ]{yangcp@hznu.edu.cn}
\author{Qi-Cheng Wu$^{5}$}
\author{Dongning Zheng$^{2,3}$}
\author{Heng Fan$^{2,3}$}
\author{Shi-Biao Zheng$^{1}$}
\email[Corresponding author, e-mail: ]{t96034@fzu.edu.cn}
\affiliation{1.Fujian Key Laboratory of Quantum Information and Quantum Optics, College of Physics and Information Engineering, Fuzhou University, Fuzhou, Fujian 350108, China}
\affiliation{2.Institute of Physics, Chinese Academy of Sciences, Beijing 100190,China}
\affiliation{3.CAS Center for Excellence in Topological Quantum Computation, Beijing 100190, China}
\affiliation{4.School of Physics, Hangzhou Normal University, Hangzhou 311121, China}
\affiliation{5.Quantum Information Research Center, Shangrao Normal University, Shangrao 334001, China}


\begin{abstract}
	Quantum entanglement between an interfering particle and a detector for
	acquiring the which-path information plays a central role for enforcing
	Bohr's complementarity principle. However, the quantitative relation between this entanglement and the fringe visibility remains untouched upon for an initial mixed state. Here we find an equality for
	quantifying this relation. Our equality characterizes how well the
	interference pattern can be preserved when an interfering particle,
	initially carrying a definite amount of coherence, is entangled, to a certain degree, with a which-path detector. This equality provides a
	connection between entanglement and interference in the unified framework of
	coherence, revealing the quantitative entanglement-interference
	complementarity. We experimentally demonstrate this relation with a
	superconducting circuit, where a resonator serves as a which-path detector
	for an interfering qubit. The measured fringe visibility of the qubit's
	Ramsey signal and the qubit-resonator entanglement exhibit a complementary
	relation, in well agreement with the theoretical prediction.
\end{abstract}


\maketitle

\section{Introduction}
Quantum coherence not only represents one of the most striking features of
quantum mechanics, but also serves as a resource for quantum information
processing \cite{Baumgratz_prl2014,Streltsov_prl2015,Chuan_Tan_pra2018}. A quantum mechanical object can simultaneously follow
different paths in either ordinary space or in Hilbert space spanned by
state vectors. Interference occurs when these paths are recombined and no
path information is available. It is impossible to obtain the particle's
path information without at the same time destroying the coherence between
the paths and the resulting interference pattern; that is, any effort to
determine which path the particle takes would involve coupling the particle
to the which-path detector (WPD) and hence correlate their states. In the
recoiling-slit gedanken experiment introduced by Einstein and Bohr \cite{Bohr_1984},
though the loss of the interference pattern is usually attributed to the
position-momentum uncertainty relation, it can also be explained in terms of
the entanglement between the photon and the two-slit apparatus induced by the
path-dependent recoil \cite{Bertet_nature2001}. In the gedanken experiment proposed by Scully et al. \cite{Scully_nature1991,Englert_apb1992}, where two high-quality micromaser cavities, each on one path of an
atomic beam passing through a two-slit assembly, act as the WPD, the atomic
path is marked by the photon the atom deposits in the corresponding cavity.
In this case, the loss of the interference pattern of the atoms has no
relation with the position-momentum uncertainty relation; instead, the
entanglement between the atomic path and the photonic state of two cavities
leads to the loss of the quantum coherence between the atomic paths, and
hence destroys the interference. The disappearance of interference due to
the introduction of a WPD has been demonstrated in different experiments
\cite{Bertet_nature2001,Scully_nature1991,Englert_apb1992,Gerry_pra1996,Zheng_oc2000,Buks_nature1998,Durr_nature1998,Durr_prl1998,Herzog_prl1995,Kim_prl2000,Liu_sciadv2017}, which cannot be explained by Heisenberg's uncertainty relation in any form.

\begin{figure*}[htbp] 
	\centering
	\includegraphics[width=6in]{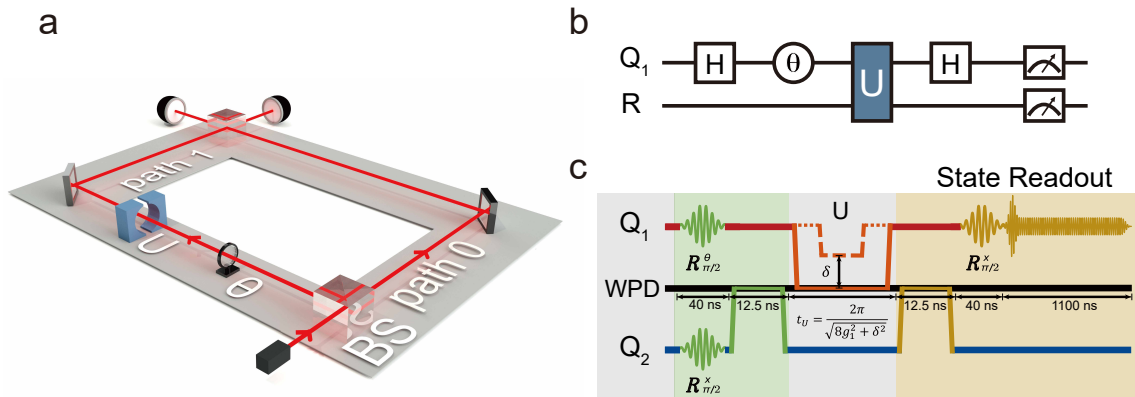}
	\caption{\textbf{Interferometer and pulse sequence.} \textbf{a} Diagram of the two-way interferometer supplemented by a WPD. A quantum system is evolved to a superposition of taking paths 0 and 1 by the first beam-splitter (BS). Then it is subjected to an adjustable relative phase shift ($\theta$), and coupled to the WPD which undergoes a unitary transformation, $U$, conditional on the system's path 1. After recombination of the two paths by the second BS, the system is detected. \textbf{b} Logic diagram of the on-chip Ramsey interferometer. A superconducting Xmon qubit, $Q_1$, acts as the interfering qubit, whose quantum state is split and recombined by two Hadamard transformations ($H$), in between which the which-path information in the quantum state space is controllably encoded on the photonic field of a resonator, $R$. A second Xmon qubit $Q_2$ is used for preparing and detecting the state of the resonator. \textbf{c} Pulse sequence. The interference experiment with which-path information acquisition consists of three parts: Preparation of the resonator's initial state $\vert W_0\rangle$ with $Q_2$; Ramsey interference supplemented with the WPD, realized by sandwiching the $Q_1$-resonator interaction between two $\pi/2$ pulses, $R_{\pi/2}^{\theta}$ and $R_{\pi/2}^{x}$; State readout.}
	\label{Fig1}
\end{figure*} 

In contrast to the case that the WPD states correlated with the two paths of
the interfering particles are completely distinguishable, the duality
relation is much more subtle in the intermediate regime \cite{Wooters_prd1979,Jaeger_pra1995,Englert_prl1996}, where only
partial which-path information is available, so that the coherence between
the paths is not completely lost and interference fringes with a reduced
visibility are retained. In Ref. \cite{Englert_prl1996}, Englert performed a detailed analysis
on this case by introducing a WPD, whose states associated with the
interfering paths are not orthogonal. He showed that, for a given amount of
which-path information stored in the WPD, quantified by the
distinguishability, $D$, the fringe visibility, $V$, is limited by an
inequality $D^{2}+V^{2}\leq 1$, which corresponds to a quantitative
description of the wave-particle duality. For an imbalanced two-way
interferometer without WPD, the inequality becomes $P^{2}+V^{2}\leq 1$,
where $P$ represents the path predictability. Recently, Bagan et al. generalized
this result to the case with multiple interfering paths, deriving the
relation between the coherence among these paths and the amount of path
information \cite{Emilio_prl2016}. The reduction of fringe visibility due to a partial
acquisition of which-path information has been observed in the optical
system \cite{Zou_prl1991}, cavity QED \cite{Bertet_nature2001,Brune_prl1996}, and superconducting circuit platform \cite{Liu_sciadv2017}.
In the atomic interference experiment reported in Ref. \cite{Durr_nature1998}, where the
internal degree of freedom of an atom acts as the WPD for its spatial paths,
the relation for the path information and fringe visibility was verified in
the intermediate regime. In this experimental demonstration, the system-WPD entanglement plays a critical role for enforcing the complementarity principle, though it was not quantified in the experiment.

In recent years, the quantitative triality relation among the visibility,
predictability, and entanglement was detailedly investigated in composite
systems that are in a pure state \cite{Jakob_pra2007,Jakob_oc2010,Tabish_ol2021,Abhinash_pra2022,Zela_ol2018,Qian_prr_2020,Yoon_SciAdv_2021,Marcos_pla2021,Qin_npjqi2019}, and was confirmed in classical optical interference experiments \cite{Qian_optica2018,Norrman_optica2020}, as well as in
single-photon experiments \cite{Qian_prr2020,Chen_npjqi2022}. An experimental investigation involving two entangled superconducting qubits has also been reported \cite{Schwaller_pra2021}, but where
    no interference experiment was performed. Here we investigate the relation
	among the fringe visibility of an interfering qubit, its quantum
	entanglement with a WPD, and the original coherence. We show that the
	system-WPD entanglement measured by concurrence and the available fringe
	visibility obey the equality $E^{2}+V^{2}=C_{0}^{2}$, where $C_{0}$ is the original coherence of the interfering system, which serves as
	a resource for producing entanglement and interference pattern. Our equality
	quantitatively characterizes the entanglement-interference complementarity
	in terms of the resource theory of coherence, in contrast with previous
	investigations. Due to the inevitable environmentally-induced decoherence effects, it is experimentally challenging to prepare a real quantum system in a perfect pure state, so that investigation of a duality relation with a
	non-unity coherence resource is of experimental relevance. We experimentally
	investigate this relation with a superconducting Ramsey interferometer,
	where a resonator serves as the WPD for a qubit in its quantum state space.
	The entanglement between the WPD and the qubit is controlled via their
	effective interaction time, achieved by the qubit's frequency tunability.
	The values of $C_0$ and $V$ are measured independently, with the
	experimental results well agreeing with theoretical predictions and
	revealing that coherence serves as a resource for both entanglement and
	interference. Unlike the photonic experiments \cite{Qian_prr2020,Chen_npjqi2022}, the present interference experiment is performed in a deterministic way, which pushes forward the experimental exploration of the entanglement-interference complementarity.  

\section{Results}
\subsection{Theoretical predictions}

Let us illustrate the underlying physics with a Mach-Zehnder interferometer, 
with the diagram shown in Fig. \ref{Fig1}a, where the interfering system, initially traveling along
path $0$, passes through a beam-splitter (BS) which splits the path $0$ into two paths $0$ and $1$. When traveling along path 1,
the system is subsequently subjected to a phase shift $\theta $ and coupled
to a WPD, which undergoes a unitary transformation, denoted as $U$. The second BS
recombines the two paths. It is convenient to denote these two paths as the
basis states $\left\vert 0\right\rangle $ and $\left\vert 1\right\rangle $,
respectively. With this notation, the system can be described as a qubit,
and the effect of each BS corresponds to the Hadamard transformation $\left\{
\left\vert 0\right\rangle \rightarrow \left( \left\vert 0\right\rangle
-{\rm i}\left\vert 1\right\rangle \right) /\sqrt{2};\text{ }\left\vert
1\right\rangle \rightarrow \left( \left\vert 1\right\rangle -{\rm i}\left\vert
0\right\rangle \right) /\sqrt{2}\right\} $. When the WPD is initially in a
pure state $\left\vert W_{0}\right\rangle $, the qubit-WPD state after their coupling is%
\begin{equation}
\left\vert \psi \right\rangle =\frac{1}{\sqrt{2}}\left( \left\vert
0\right\rangle \left\vert W_{0}\right\rangle -{\rm ie}^{{\rm i}\theta }\left\vert
1\right\rangle U\left\vert W_{0}\right\rangle \right) .
\end{equation}%
After the second Hadamard transformation, the probability of detecting the
qubit in the state $\left\vert 1\right\rangle $ is
\begin{equation}\label{e2}
P_{1}=\frac{1}{2}\left[ 1+V_{0}\cos \left( \theta +\phi \right) \right] ,
\end{equation}%
where $V_{0}=\left\vert \left\langle W_{0}\right\vert U\left\vert
W_{0}\right\rangle \right\vert $ and $\phi =\arg \left( \left\langle
W_{0}\right\vert U\left\vert W_{0}\right\rangle \right) $. This $\theta $%
-dependent probability manifests the interference behavior, with the fringe
visibility reduced to $V_{0}$ due to the qubit-WPD entanglement. When the
WPD lies in a multi-dimensional Hilbert space, we map it to the
two-dimensional space \{$\left\vert W_{0}\right\rangle ,\left\vert
W_{1}\right\rangle $\} so that the qubit-WPD entanglement can be measured
with concurrence \cite{Wootters_prl1998}, where
\begin{equation}
\left\vert W_{1}\right\rangle =\sqrt{\frac{1}{1-V_{0}^{2}}}\left(
U\left\vert W_{0}\right\rangle -V_{0}{\rm e}^{{\rm i}\phi }\left\vert W_{0}\right\rangle
\right)
\end{equation}%
is orthogonal to $\left\vert W_{0}\right\rangle $. With this mapping, the
joint qubit-WPD state can be considered as a two-qubit entangled state. The concurrence for
measuring the amount of the qubit-WPD entanglement is $E=\sqrt{1-V_{0}^{2}}$.
This concurrence is equivalent to the distinguishability, $D$, which
measures the path information stored in the WPD, and is defined as
\begin{equation}
D=\frac{1}{2}{\rm Tr}_{{\rm w}}\left\vert \rho _{{\rm w},0}-\rho _{{\rm w},1}\right\vert ,
\end{equation}
where $\rho _{{\rm w},0}=\left\langle 0\right\vert \rho \left\vert 0\right\rangle /
{\rm tr}_{{\rm w}}\left( \left\langle 0\right\vert \rho \left\vert 0\right\rangle
\right) $ and $\rho _{{\rm w},1}=\left\langle 1\right\vert \rho \left\vert
1\right\rangle /{\rm tr}_{{\rm w}}\left( \left\langle 1\right\vert \rho \left\vert
1\right\rangle \right) $ respectively denote the density operators of the
WPD associated with the interfering qubit's $\left\vert 0\right\rangle $ and
$\left\vert 1\right\rangle $ states after their coupling, with $\rho
=\left\vert \psi \right\rangle \left\langle \psi \right\vert $ denoting the
density operator for the joint qubit-WPD state, and the tracing is over the
WPD's degree of freedom.

We note that $E=D$ holds only when the two paths of the interfering system
are completely coherent. When the coherence between the two paths is
partially lost before the qubit-WPD coupling, the corresponding qubit's
state is described by the density operator
\begin{widetext}
\begin{equation}\label{e7}
\rho _{{\rm q},0}=\frac{1}{2}\left[ \left\vert 0\right\rangle \left\langle
0\right\vert +\left\vert 1\right\rangle \left\langle 1\right\vert
+{\rm i}C_{0}\left( {\rm e}^{-{\rm i}\theta }\left\vert 0\right\rangle \left\langle
1\right\vert -{\rm e}^{{\rm i}\theta }\left\vert 1\right\rangle \left\langle
0\right\vert \right) \right] ,
\end{equation}%
\end{widetext}
where $C_{0}$ is the coherence between the two paths, defined as the sum of
the moduli of the off-diagonal elements \cite{Baumgratz_prl2014}.

After the qubit-resonator coupling and the subsequent Hadamard transformation, the probability of detecting the
qubit in the state $\left\vert 1\right\rangle $ is
\begin{equation}
P_{1}^{\prime}=\frac{1}{2}\left[ 1+V\cos \left( \theta +\phi \right) \right] ,
\end{equation}%
where $V=C_{0}V_{0}$ is the fringe visibility. This visibility and the qubit-WPD entanglement satisfy the equality (see Supplementary Note 2)
\begin{equation}\label{new eq7}
	E^{2}+V^{2}=C_{0}^{2}.
\end{equation}%
This implies that both the qubit-WPD concurrence and the fringe visibility are reduced by a factor of $C_0$ due to the decoherence between the paths before the qubit-WPD coupling. In distinct contrast, the distinguishability, $D=\sqrt{1-V_0^2}$, is not affected by the decoherence, and consequently,
$D^{2}+V^{2}>C_{0}^{2}$ unless $C_{0}=1$. The results can
be interpreted as follows. The conditional unitary evolution arising from
the qubit-WPD coupling corresponds to an incoherent operation, which cannot
produce entanglement when the qubit is in an incoherent state before this
operation \cite{Chuan_Tan_pra2018}. Both the interference and the qubit-WPD entanglement
originate from the original coherence of the interfering qubit; this
coherence, as a resource, can be either converted to entanglement or used
for interference. For a given coherence resource, the more the amount of
entanglement, the weaker the interference effect. In contrast, the
distinguishability does not depend on the coherence resource; instead, it is
determined by the distance between the WPD states associated with the two
paths. For example, even for $C_{0}=0$, $D$ can reach 1 as long as the two
WPD states are orthogonal. The equality of Eq. (\ref{new eq7}) also holds for the unbalanced interferometer, where the two paths are not equally populated
before coupling to the WPD (see Supplementary Note 1). We note that this equality represents a more general triality relation, which reduces to the previously proposed triality equality \cite{Jakob_pra2007,Jakob_oc2010,Tabish_ol2021,Abhinash_pra2022,Zela_ol2018,Qian_prr_2020,Yoon_SciAdv_2021,Marcos_pla2021,Qin_npjqi2019} when the interfering qubit is initially a pure state so that $C_{0}^{2}=1-P^{2}$, where $P$ denotes the predictability.

\begin{table}
	\centering
	\renewcommand{\arraystretch}{1.5}
	\begin{tabular}{{ccc}}
		\hline\hline
		\centering
		Parameters & \qquad\qquad $Q_{1}$ &\qquad\qquad  $Q_{2}$  \qquad\qquad \qquad   \\ \hline
		\centering
		$\omega_{j}/2\pi$ &\qquad\qquad  $\SI{5.967}{\giga\hertz}$    &  $\SI{5.354}{\giga\hertz}$ \\
		
		$g_{j}/2\pi$&\qquad\qquad  $\SI{19.2}{\mega\hertz} $    &  $\SI{19.9}{\mega\hertz}$     \\
		
		$T_{1,j}$  &\qquad\qquad  $17.1$  $\SI{}{\micro\second}$   &$23.4$ $\SI{}{\micro\second}$   \\
		
		$T_{\varphi,j}$   &\qquad\qquad  $3.6$  $\SI{}{\micro\second}$   &$2.7$  $\SI{}{\micro\second}$     \\  \hline\hline
		
	\end{tabular}
	\caption{\label{table1}Experimental parameters. $\omega_j/2\pi$: $Q_j$'s idle frequency; $g_{j}$: $Q_j$-$R$ coupling strength; $T_{1,j}$: $Q_j$'s energy relaxation time; $T_{\varphi,j}$: $Q_j$'s pure dephasing time. The other parameters can be found in the Supplementary Table 1.} 
	
\end{table}

\subsection{Device and experimental scheme}
These theoretical predictions can be demonstrated on different platforms with controlled natural or artificial atoms \cite{Buluta_ropip2011,Geogescu_rmp2014}, among which circuit quantum electrodynamics systems represent a typical example owing to the available strong coupling between superconducting qubits and microwave photons \cite{You_nature2011,Gu_pr2017}. The experiment is implemented with a circuit Ramsey interferometer, with the logic diagram of the process sketched in Fig. \ref{Fig1}b. The device involves 5 frequency-tunable Xmon qubits \cite{song_nc2017,Ning_prl2019,Yang_npjqi2021,Xu_optica2021}, one of which ($Q_1$) acts as the interfering qubit, whose two lowest levels, denoted as $\left\vert 0\right\rangle $ and $\left\vert 1\right\rangle $, correspond to two interfering paths in the Hilbert space. Such a Ramsey interferometer is analogous to the Mach-Zehnder interferometer, where the BSs are replaced by two Hadamard transformations ($H$). The photonic field stored in a bus resonator, $R$, is used to acquire the which-path information of $Q_1$ before the recombination of the two interfering paths. A second qubit $Q_2$ serves as the ancilla for preparing and reading out the state of the resonator. In our experiment, the tunable phase shift $\theta$ between the two interfering paths is incorporated into the first microwave pulse. The experimental parameters are detailed in Table 1.

To enable the resonator to act as the WPD, before coupling to $Q_{1}$ it is
prepared in the coherent superposition containing $0$ and $1$ photon: $%
\left\vert W_{0}\right\rangle =(\left\vert 0_{{\rm r}}\right\rangle -\left\vert
1_{{\rm r}}\right\rangle )/\sqrt{2}$. The resonator's conditional unitary
evolution, $U$, is realized by tuning the frequency associated with $Q_{1}$%
's transition $\left\vert 1\right\rangle \leftrightarrow \left\vert
2\right\rangle $, $\omega _{12}$, close to the resonator's frequency, where
$\left\vert 2\right\rangle $ represents the second excited state of $Q_{1}$.
After an interaction time $\tau =\pi /\Omega $, the resonator undergoes an
evolution, $U={\rm e}^{{\rm i}\beta \left\vert 1_{\rm r}\right\rangle \left\langle
	1_{\rm r}\right\vert }$, conditional on $Q_{1}$'s state $\left\vert
1\right\rangle $, where $\beta =\pi \lbrack 1-\delta /(2\Omega)]$, with $%
\Omega =\sqrt{2g_{1}^{2}+\delta ^{2}/4}$ and $\delta =\omega _{12}-\omega
_{\rm r}$ \cite{Rauschenbeutel_prl1999,Mariantoni_science2011}. This conditional dynamics correlates the two paths (the qubit
being in $\left\vert 0\right\rangle $ or $\left\vert 1\right\rangle $)
produced by the first $\pi /2$ pulse ($R_{\pi /2}^{\theta }$) with the
resonator's states $\left\vert W_{0}\right\rangle $ and $U\left\vert
W_{0}\right\rangle $, respectively. After the second $\pi /2$ pulse $R_{\pi /2}^{x}$, the
probability of detecting the qubit in $\left\vert 1\right\rangle $ state is
given by Eq. (\ref{e2}), which corresponds to a $\theta $-dependent Ramsey
interference signal, with the fringe visibility $V_{0}=\cos (\beta /2)$.

\subsection{Measurement of the entanglement-interference relation for a unity source coherence}
\begin{figure*}[t] 
	\centering
	\includegraphics[width=7.0in]{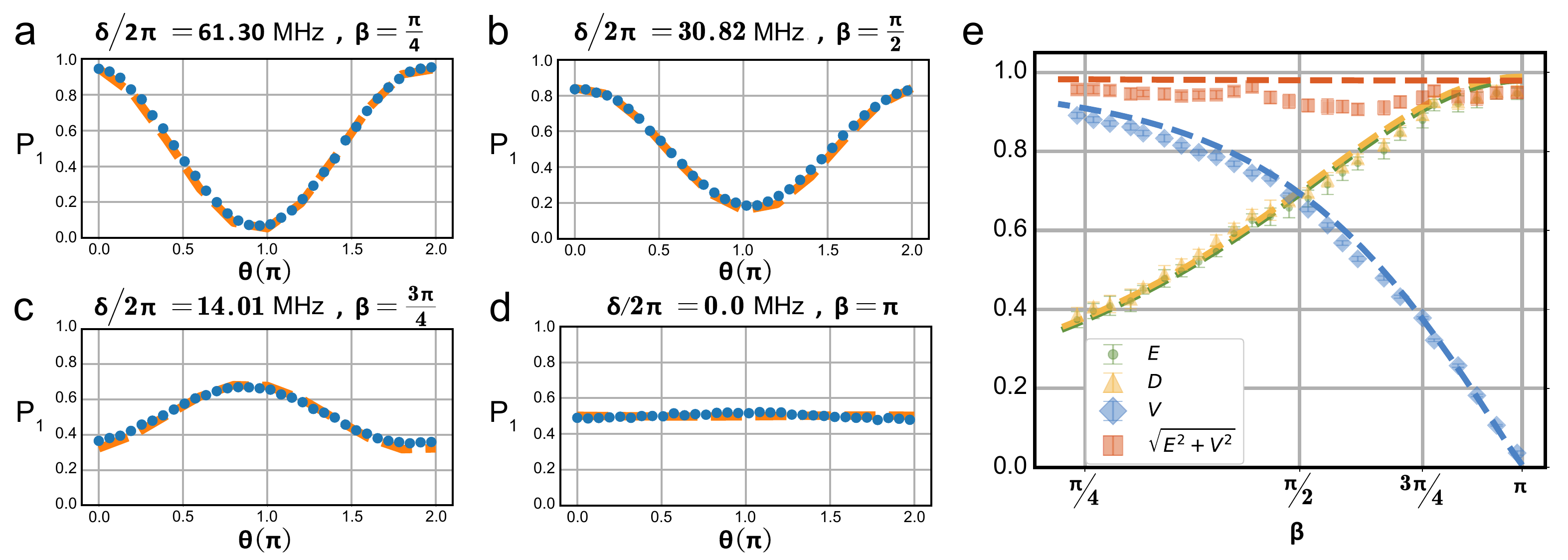}
	\caption{\textbf{Measured entanglement-interference relation for a unity source coherence.} The Ramsey interference is revealed by the probability of measuring the test qubit in the state $\left\vert 1\right\rangle $ as a function of $\theta $ for different conditional phase shift $\beta$:
		\textbf{a} $\pi/4$; \textbf{b} $\pi/2$; \textbf{c} $3\pi/4$; \textbf{d} $\pi$. 
		The corresponding fringes visibilities are $0.8853$, $0.6473$, $0.3184$, and $0.0311$, respectively. \textbf{e} Measured fringe visibility $V$
		(diamonds), $Q_{1}$-$Q_{2}$ concurrence $E$ (dots), and distinguishability $ D $ (triangles) as a function of the conditional phase shift $\beta $ of the resonator. $E$ is measured after mapping the resonator's state back to $	Q_{2} $. The results are obtained with the pulse sequences shown in the Supplementary Fig. 5.
		Squares denote the quantity $\sqrt{E^{2}+V^{2}}$. The dashed lines correspond to numerical results. The error bars are the standard deviation of the measurements. For some data points, the error bar is smaller than the marker.}
	\label{Fig2}
\end{figure*}

The interference
experiment starts by simultaneously driving the ancilla qubit $Q_{2}$ from the ground
state $\left\vert 0_{{\rm a}}\right\rangle $ to the superposition state $%
(\left\vert 0_{{\rm a}}\right\rangle -{\rm i}\left\vert 1_{\rm a}\right\rangle )/\sqrt{2}$
with a $\pi /2$ pulse $R_{\pi /2}^{x}$, and applying a $\pi/2$ pulse $R_{\pi /2}^{\theta }$ to $Q_1$ (Fig. \ref{Fig1}c), which produces a $\pi/2$ rotation around the axis with an angle $\theta$ to $x-$axis on the equatorial plane of the Bloch sphere. Then $Q_{2}$ is tuned on-resonance with
the resonator initially in the vacuum state $\left\vert 0_{\rm r}\right\rangle $ for
a time $\SI{12.5}{\nano\second}$, which realizes a swapping operation, mapping the ancilla state
to the resonator, preparing it in the superposition of the zero- and
one-photon states required for storing the which-path information of the
interfering qubit. After preparation of the WPD's state $\left\vert
W_{0}\right\rangle $, $Q_{2}$ is biased back to its idle frequency.
Subsequently, $Q_{1}$'s transition $\left\vert 1\right\rangle
\leftrightarrow \left\vert 2\right\rangle $ is tuned close to the
resonator's frequency to realize the conditional evolution $U$, resulting in entanglement between $Q_{1}$ and the resonator. Following this
conditional dynamics, $Q_{1}$ is tuned back to its idle frequency, where the
second $\pi /2$ pulse $R_{\pi /2}^{x}$ is applied.
The Ramsey signals, defined as the probability for measuring $Q_{1}$ in $\left\vert
1\right\rangle $ state as a function of $\theta $, for conditional phase
shifts $\beta =\pi/4$, $\pi/2$, $3\pi/4$, and $\pi$ are displayed
in Fig. \ref{Fig2}a, b, c, d, respectively. As expected, the fringe
visibility decreases as $\beta$ increases for $0\leq \beta \leq \pi $, as a
result of the increasing entanglement between $Q_{1}$ and the resonator. The
extracted fringe visibility as a function of $\beta $ is denoted by diamonds
in Fig. \ref{Fig2}e, which is in well agreement with the numerical simulation (blue dashed line). We note that the qubit has a slight probability of being populated in the $\left\vert 2\right\rangle$-state after the
conditional phase shift, as a consequence of experimental imperfections. In
the experiment, the maximal residual population of this state is measured to
be $0.022$, which only changes the Ramsey fringe visibility by about $1.29\%$.

To quantitatively investigate the change of entanglement between $Q_{1}$ and
the resonator accompanying the decrease of the fringe visibility, we map the
state of the resonator back to $Q_{2}$ for joint state tomography. This is
achieved by tuning $Q_{2}$ on resonance with the resonator for a time of $\SI{12.5}{\nano\second}$ after the resonator's interaction with $Q_{1}$, and then biasing back it
to its idle frequency. Then the joint $Q_{1}$-$Q_{2}$ density matrix is
measured by quantum state tomography (see Supplementary Note 10). The
resulting $Q_{1}$-$Q_{2}$ concurrence as a function of $\beta $ is
represented by dots in Fig. \ref{Fig2}e, which agrees with the simulation (green dashed line). As expected, the higher the concurrence, the lower the fringe
contrast. The source coherence $C_{0}$ is measured through quantum
state tomography on $Q_{1}$ right after the first $\pi /2$ pulse $R_{\pi
	/2}^{\theta }$; preparation of the WPD state $\left\vert W_{0}\right\rangle $%
, coupling $Q_{1}$ to the WPD, and subsequent WPD-$Q_{2}$ state mapping are
unnecessary for measuring $C_{0}$. The quantity $\sqrt{E^{2}+V^{2}}$ as a
function of $\beta $ is also displayed in Fig. \ref{Fig2}e (squares), which well
coincides with the measured $C_{0}$. The slight difference between the
measured values of these quantities mainly results from energy relaxation
and dephasing of the qubits and photon loss of the resonator.

\begin{figure}[htbp] 
	\centering
	\includegraphics[width=3.4in]{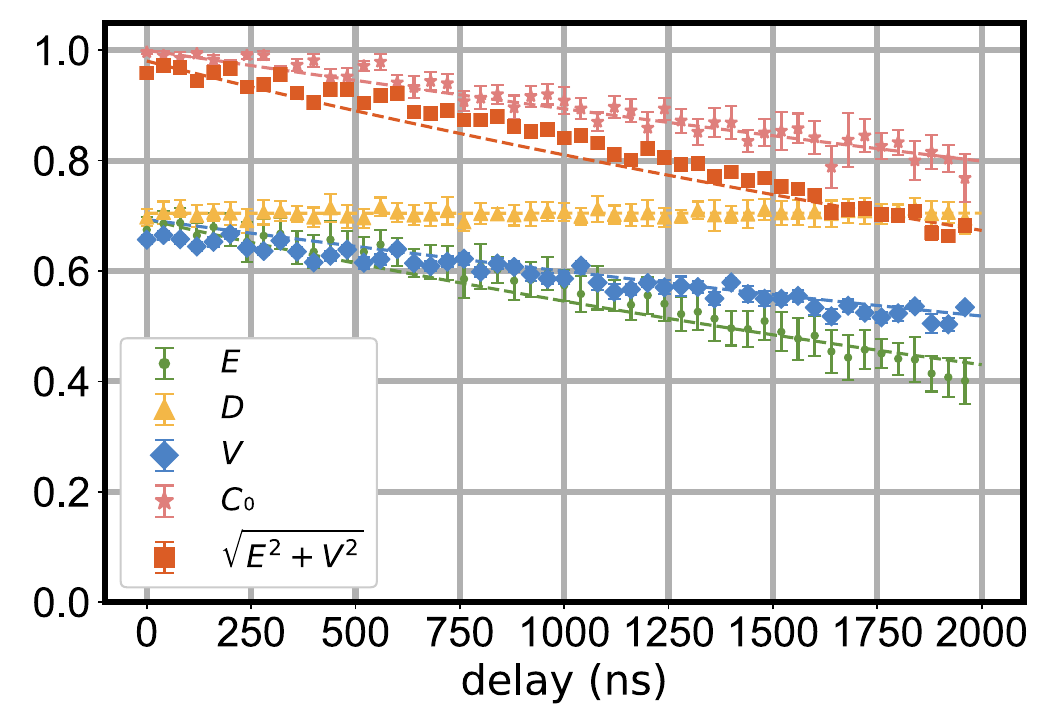}
	\caption{\textbf{Measured entanglement-interference relations for  different source coherences.} The diamonds, dots, stars, and triangles denote the fringe visibility, concurrence, source coherence, and distinguishability, respectively. The results are obtained with the preparation of the WPD's superposition state $%
		\left\vert W_{0}\right\rangle $ delayed by a preset time. The conditional phase shift of the resonator induced by coupling to $Q_{1}$ is $\beta =\pi /2$. Squares denote evolution of the quantity $\sqrt{E^{2}+V^{2}}$. The error bars represent the standard deviation of the measurements. 
		For some data points, the error bar is smaller than the corresponding symbol.}
	\label{Fig3}
\end{figure}

We further reveal the relation between the distinguishability of the two paths and
the entanglement between the interfering qubit and the WPD for the fully
coherent qubit state. This is achieved by extracting the density matrices $\rho _{{\rm w},0}$ and $%
\rho _{{\rm w},1}$ of the WPD associated with the interfering qubit's $\left\vert
0\right\rangle $ and $\left\vert 1\right\rangle $ states after their
coupling, obtained by projecting the measured $Q_{1}$-$Q_{2}$ density matrix
$\rho $ to $Q_{1}$'s basis states $\left\vert 0\right\rangle $ and $%
\left\vert 1\right\rangle $, which are displayed in Supplementary Note 3.
The resulting distinguishability $D$, as a function of the conditional phase $%
\beta $, is denoted by the triangles of Fig. \ref{Fig2}e, which coincides with the $Q_{1}$-%
$Q_{2}$ concurrence (green dots) very well, verifying the WPD acquires the
which-path information by entangling its state with the interfering qubit.

\subsection{Measurement of entanglement-interference relations for different source coherences}
To further confirm the quantitative relation among the system's original
coherence $C_{0}$, fringe visibility $V$, and the system-WPD entanglement $E$
for partially coherent qubit states, we adjust the temporal orders of the $%
\pi /2$ pulse ($R_{\pi /2}^{\theta }$) of the Ramsey interferometer and the
preparation of the WPD's superposition state $\left\vert W_{0}\right\rangle $%
, and delay this preparation procedure for a time $t$ after $R_{\pi
	/2}^{\theta }$. Due to the energy relaxation and dephasing during this
delay, the density operator of $Q_{1}$ is given by Eq. (\ref{e7}), with $%
C_{0}\simeq {\rm e}^{-t/2T_{1}-t/T_{\varphi}}$. The conditional phase shift of
the resonator induced by coupling to $Q_{1}$ is set to $\beta =\pi /2$. The
measured fringe visibility $V$ (diamonds), $Q_{1}$-$Q_{2}$ concurrence $E$
(dots), the system's original coherence $C_{0}$ (stars), and the
distinguishability $D$ (triangles), as functions of the delay $t$, are displayed
in Fig. \ref{Fig3}. As expected, both $V$ and $E$ decrease as the delay increases.
In addition to the decoherence of the test qubit, the concurrence is affected by thermal excitations of the ancilla and the resonator during the delay, which accounts for the result that the measured concurrence decays faster than the visibility, so that the measured $\sqrt{E^{2}+V^{2}}$ (squares) is smaller than $C_0$. On the other hand, measured $D$ (triangles)
almost does not change with the delay, which implies that it is independent
of the source coherence.

\section{Discussion}
We have derived an equality relating the fringe visibility of an interfering
system and its entanglement with a WPD inside an interferometer to the
original coherence. The result reveals, for a given amount of coherence
resource, the available fringe visibility is determined by the system-WPD
entanglement: the stronger the system's entanglement with the WPD, the lower
the available fringe visibility. We perform an experimental test of this
relation in a superconducting circuit, in which a resonator is prepared in a
superposition of zero- and one-photon states and acts as the WPD for a
superconducting Xmon qubit. The measured results agree well with the theoretical
predictions, confirming the entanglement-interference complementarity.

\section{Data Availability}
All data needed to evaluate the conclusions in the paper are present in the paper
and/or the Supplementary Materials. Additional data related to this paper may be
requested from the authors.

\section{Acknowledgments}
This work was supported by the National Natural
Science Foundation of China (Grant No. 12274080,
No. 11875108, No. 11934018, No. 92065114, No. T2121001, No. 12264040, and No. U21A20436), Innovation Program for Quantum Science and Technology (Grant No. 2021ZD0300200), the Strategic Priority Research Program of Chinese Academy of Sciences (Grant
No. XDB28000000), the Key-Area Research and Development Program of Guangdong Province, China (Grant No. 2020B0303030001), Beijing Natural Science Foundation (Grant No. Z200009), and Project from Fuzhou University under (Grant No. JG202001-2, No. 049050011050).

\section{Author Contributions}
S.-B.Z. derived the equality and conceived the experiment. X.-J.H. and
P.-R.H. performed the experiment and analyzed the data, under supervision of
Z.-B.Y. and S.-B.Z. S.-B.Z., Z.-B.Y., and C.-P.Y. cowrote the manuscript
with feedbacks from all authors. All authors contributed to interpretation
of the observed phenomena and helped to improve presentation of the
manuscript.

\section{Competing Interests}
The authors declare that they have no competing interests.


\begin{references}
\bibitem{Baumgratz_prl2014} Baumgratz, T., Cramer, M. \& Plenio, M. B. Quantifying coherence. {\it Phys. Rev. Lett.} {\bf 113,} 140401 (2014). 

\bibitem{Streltsov_prl2015} Streltsov, A., Singh, U., Dhar, H. S., Bera, M. N. \& Adesso, G. Measuring quantum coherence with entanglement. {\it Phys. Rev. Lett.} {\bf 115,} 020403 (2015). 

\bibitem{Chuan_Tan_pra2018} Tan, K. C., Choi, S., Kwon, H. \& Jeong, H. Coherence, Quantum Fisher information, superradiance, and entanglement as interconvertible resources. {\it Phys. Rev. A} {\bf 97,} 052304 (2018) 

\bibitem{Bohr_1984} Bohr, N. {\it Quantum Theory and Measurement} (eds Wheeler, J.
A. \& Zurek, W. H.) 9-49 (Princeton University Press, Princeton, NJ, 1984). 

\bibitem{Bertet_nature2001} Bertet, P. et al. A complementarity experiment with an interferometer at the quantum-classical boundary. {\it Nature} {\bf 411,} 166-170 (2001).

\bibitem{Scully_nature1991} Scully, M. O., Englert, B.-G. \& Walther, H. Quantum optical tests of complementarity. {\it Nature} {\bf 351,} 111-116 (1991). 

\bibitem{Englert_apb1992} Englert, B.-G., Walther, H. \& Scully, M. O. Quantum optical ramsey fringes and complementarity. {\it Appl. Phys. B} {\bf 54,} 366-368 (1992). 


\bibitem{Gerry_pra1996} Gerry, C. C. Complementarity and quantum erasure with
dispersive atom-field interactions. {\it Phys. Rev. A} {\bf 53,} 1179-1182 (1996). 

\bibitem{Zheng_oc2000} Zheng, S.-B. A simplified scheme for testing complementarity and realizing quantum eraser. {\it Opt. Commun.} {\bf 173,} 265-267 (2000).

\bibitem{Buks_nature1998} Buks, E., Schuster, R., Heiblum, M., Mahalu, D. \& Umansky, V. Dephasing in electron interference by a `which-path' detector. {\it Nature} {\bf 391,} 871-874 (1998). 

\bibitem{Durr_nature1998} D\"{u}rr, S., Nonn, T. \& Rempe, G. Origin of quantum-mechanical complementarity probed by a `which-way' experiment in an atom interferometer. {\it Nature} {\bf 395,} 33-37 (1998). 

\bibitem{Durr_prl1998} D\"urr, S., Nonn, T. \& Rempe, G. Fringe visibility and which-way information in an atom interferometer. {\it Phys. Rev. Lett.} {\bf 81,} 5705-5709 (1998). 

\bibitem{Herzog_prl1995} Herzog, T. J., Kwiat, P. G., Weinfurter, H. \& Zeilinger, A.
Complementarity and the quantum eraser. {\it Phys. Rev. Lett.} {\bf 75,} 3034-3037 (1995).

\bibitem{Kim_prl2000} Kim, Y.-H., Yu, R., Kulik, S. P., Shih, Y. \& Scully, M. O. Delayed \lq\lq choice\rq\rq quantum eraser. {\it Phys. Rev. Lett.} {\bf 84,} 1-5 (2000).

\bibitem{Liu_sciadv2017} Liu, K. et al. A twofold quantum delayed-choice experiment in a superconducting circuit. {\it Sci. Adv.} {\bf 3,}
e1603159 (2017). 

\bibitem{Wooters_prd1979} Wootters, W. K. \& Zurek, W. H. Complementarity in the double-slit experiment: quantum nonseparability and a quantitative statement of Bohr's principle. {\it Phys. Rev. D} {\bf 19,} 473-484 (1979). 

\bibitem{Jaeger_pra1995} Jaeger, G., Shimony, A. \& Vaidman, L. Two interferometric complementarities. {\it Phys. Rev. A} {\bf 51,} 54-67 (1995). 

\bibitem{Englert_prl1996} Englert, B.-G. Fringe visibility and which-way information: an inequality. {\it Phys. Rev. Lett.} {\bf 77,} 2154-2157 (1996). 

\bibitem{Emilio_prl2016} Bagan, E., Bergou, J. A., Cottrell, S. S. \& Hillery, M. Relations between coherence and path information. {\it Phys. Rev. Lett.} {\bf 116,} 160406 (2016). 

\bibitem{Zou_prl1991} Zou, X. Y., Wang, L. J. \& Mandel, L. Induced coherence and indistinguishability in optical interference. {\it Phys. Rev. Lett.} {\bf 67,} 318-321 (1991). 

\bibitem{Brune_prl1996} Brune, M. et al. Observing the progressive
decoherence of the \textquotedblleft meter\textquotedblright\ in a quantum
measurement. {\it Phys. Rev. Lett.} {\bf 77,} 4887-4890 (1996).

\bibitem{Jakob_pra2007} Jakob, M. \& Bergou, J. A. Complementarity and entanglement in bipartite qudit systems. {\it Phys. Rev. A} {\bf 76,} 052107 (2007). 

\bibitem{Jakob_oc2010} Jakob, M. \& Bergou, J. A. Quantitative complementarity relations in bipartite systems: entanglement as a physical reality. {\it Opt. Commun.} {\bf 283,} 827-830 (2010). 

\bibitem{Tabish_ol2021} Qureshi, T. Predictability, distinguishability, and
entanglement. {\it Opt. Lett.} {\bf 46,} 492-495 (2021). 

\bibitem{Abhinash_pra2022} Roy, A. K., Pathania, N., Chandra, N. K., Panigrahi, P. K. \& Qureshi, T. Coherence, path predictability, and {\it I}
concurrence: a triality. {\it Phys. Rev. A} {\bf 105,} 032209 (2022). 

\bibitem{Zela_ol2018} De Zela, F. Optical approach to concurrence and polarization.
{\it Opt. Lett.} {\bf 43,} 2603-2606 (2018).

\bibitem{Qian_prr_2020} Qian, X.-F. \& Agarwal, G. S. Quantum duality: a source point of view. {\it Phys. Rev. Res.} {\bf 2,} 012031(R) (2020). 

\bibitem{Yoon_SciAdv_2021} Yoon, T. H. \& Cho, M. Quantitative complementarity of wave-particle duality. {\it Sci. Adv.} {\bf 7,} eabi9268 (2021). 

\bibitem{Marcos_pla2021} Basso, M. L. W. \& Maziero, J. Entanglement monotones
connect distinguishability and predictability. {\it Phys. Lett. A} {\bf 425,} 127875
(2021). 

\bibitem{Qin_npjqi2019} Qin, W., Miranowicz, A., Long, G., You, J. Q. \& Nori, F. Proposal to test quantum wave-particle superposition on massive mechanical resonators.
{\it npj Quantum Inf.} {\bf 5,} 58 (2019). 

\bibitem{Qian_optica2018} Qian, X.-F., Vamivakas, A. N. \& Eberly, J. H. Entanglement
limits duality and vice versa. {\it Optica} {\bf 5,} 942-947 (2018). 

\bibitem{Norrman_optica2020} Norrman, A., Friberg, A. T. \& Leuchs, G. Vector-light quantum complementarity and the degree of polarization. {\it Optica} {\bf 7,} 93-97 (2020). 

\bibitem{Qian_prr2020} Qian, X.-F. et al. Turning off quantum duality. {\it Phys. Rev.
Res.} {\bf 2,} 012016(R) (2020). 

\bibitem{Chen_npjqi2022} Chen, D.-X. et al. Experimental investigation of wave-particle duality relations in asymmetric beam interference. {\it npj Quantum Inf.} {\bf 8,} 101 (2022). 

\bibitem{Schwaller_pra2021} Schwaller, N., Dupertuis, M. A. \& Javerzac-Galy, C. Evidence of the quantum entanglement constraint on wave-particle duality using the IBM Q quantum computer. {\it Phys. Rev. A} {\bf 103,} 022409 (2021).

\bibitem{Wootters_prl1998} Wootters, W. K. Entanglement of formation of an arbitrary state of two qubits. {\it Phys. Rev. Lett.} {\bf 80,} 2245-2248 (1998). 

\bibitem{Buluta_ropip2011} Buluta, I., Ashhab, S. \& Nori, F. Natural and artificial atoms for quantum computation. {\it Rep. Prog. Phys.} {\bf 74,} 104401 (2011). 

\bibitem{Geogescu_rmp2014} Georgescu, I. M., Ashhab, S. \& Nori, F. Quantum Simulation. {\it Rev. Mod. Phys.} {\bf 86,} 153-185 (2014). 

\bibitem{You_nature2011} You, J. Q. \& Nori, F. Atomic physics and quantum optics using superconducting circuits. {\it Nature} {\bf 474,} 589-597 (2011). 

\bibitem{Gu_pr2017} Gu, X., Kockum, A. F., Miranowicz, A., Liu, Y.-X. \& Nori, F.
Microwave photonics with superconducting quantum circuits. {\it Phys. Rep.}
{\bf 718-719,} 1-102 (2017). 

\bibitem{song_nc2017} Song, C. et al. Continuous-variable geometric phase and its manipulation for quantum computation in a superconducting circuit. {\it Nat. Commun.} {\bf 8,} 1061 (2017). 

\bibitem{Ning_prl2019} Ning, W. et al. Deterministic entanglement swapping in a superconducting circuit. {\it Phys. Rev. Lett.} {\bf 123,}
060502 (2019). 

\bibitem{Yang_npjqi2021} Yang, Z.-B. et al. Experimental demonstration of entanglement-enabled universal quantum cloning in a circuit. {\it npj Quantum Inf.} {\bf 7,} 44 (2021). 

\bibitem{Xu_optica2021} Xu, K. et al. Demonstration
of a non-Abelian geometric controlled-NOT gate in a superconducting circuit. {\it Optica} {\bf 8,} 972-976 (2021). 

\bibitem{Rauschenbeutel_prl1999} Rauschenbeutel, A. et al. Coherent operation of a tunable quantum phase gate in cavity QED. {\it Phys. Rev. Lett.} {\bf 83,} 5166-5169 (1999). 

\bibitem{Mariantoni_science2011} Mariantoni, M. et al. Implementing the quantum von Neumann architecture with superconducting circuits. {\it Science} {\bf 334,} 61-65 (2011). 
\end{references}
\end{document}


\title{Supplementary material for {\textquotedblleft}Entanglement-interference complementarity and experimental demonstration in a superconducting circuit{\textquotedblright}}
\author{Xin-Jie Huang$^{1}$}
\thanks{These authors contributed equally to this work.}
\author{Pei-Rong Han$^{1}$}
\thanks{These authors contributed equally to this work.}
\author{Wen Ning$^{1}$}
\author{Shou-Bang Yang$^{1}$}
\author{Xin Zhu$^{1}$}
\author{Jia-Hao L$\ddot{u}$$^{1}$}
\author{Ri-Hua Zheng$^{1}$}
\author{Hekang Li$^{2}$}
\author{Zhen-Biao Yang$^{1}$}
\email{zbyang@fzu.edu.cn}
\author{Kai Xu$^{2,3}$}
\author{Chui-Ping Yang$^{4}$}
\email{yangcp@hznu.edu.cn}
\author{Qi-Cheng Wu$^{5}$}
\author{Dongning Zheng$^{2,3}$}
\author{Heng Fan$^{2,3}$}
\author{Shi-Biao Zheng$^{1}$}
\email{t96034@fzu.edu.cn}
\affiliation{1.Fujian Key Laboratory of Quantum Information and Quantum Optics, College of Physics and Information Engineering, Fuzhou University, Fuzhou, Fujian 350108, China}
\affiliation{2.Institute of Physics, Chinese Academy of Sciences, Beijing 100190,China}
\affiliation{3.CAS Center for Excellence in Topological Quantum Computation, Beijing 100190, China}
\affiliation{4.School of Physics, Hangzhou Normal University, Hangzhou 311121, China}
\affiliation{5.Quantum Information Research Center, Shangrao Normal University, Shangrao 334001, China}
\maketitle

	\section{Supplementary Note 1: System-WPD concurrence}
	
	When the coherence between the two paths is partially lost before the
	qubit-WPD coupling due to dephasing, the joint state of this qubit and the
	WPD after their coupling can be rewritten as 
		\begin{eqnarray}
			\rho &=&\frac{1}{2}\left[ | 0\rangle \langle
			0| \otimes | W_{0}\rangle \langle
			W_{0}| +| 1\rangle \langle 1|
			\otimes \left( V_{0}{\rm e}^{{\rm i}\phi }| W_{0}\rangle +\sqrt{%
				1-V_{0}^{2}}| W_{1}\rangle \right) \left( \langle
			W_{0}| V_{0}{\rm e}^{-{\rm i}\phi }+\langle W_{1}| \sqrt{%
				1-V_{0}^{2}}\right) \right.  \nonumber\\
			&&+{\rm i}C_{0}{\rm e}^{-{\rm i}\theta }| 0\rangle \langle 1|
			\otimes | W_{0}\rangle \left( \langle W_{0}|
			V_{0}{\rm e}^{-{\rm i}\phi }+\langle W_{1}| \sqrt{1-V_{0}^{2}}\right) \nonumber \\
			&&\left. -{\rm i}C_{0}{\rm e}^{{\rm i}\theta }| 1\rangle \langle
			0| \otimes \left( V_{0}{\rm e}^{{\rm i}\phi }| W_{0}\rangle +%
			\sqrt{1-V_{0}^{2}}| W_{1}\rangle \right) \langle
			W_{0}| \right] ,
		\end{eqnarray}
	where $V_{0}=| \langle W_{0}| U|
	W_{0}\rangle | $ and $\phi =\arg \left( \langle
	W_{0}| U| W_{0}\rangle \right) $. In the joint
	qubit-WPD basis $\left\{ | 0\rangle |
	W_{0}\rangle ,| 0\rangle | W_{1}\rangle
	,| 1\rangle | W_{0}\rangle ,|
	1\rangle | W_{1}\rangle \right\} $, $\rho $ can be
	expressed in the matrix form%
		\begin{equation}
		\rho =\frac{1}{2}\left( 
		\begin{array}{cccc}
		1 & 0 & {\rm i}C_{0}V_{0}{\rm e}^{-{\rm i}(\theta +\phi )} & {\rm i}C_{0}\sqrt{1-V_{0}^{2}}%
		{\rm e}^{-{\rm i}\theta } \\ 
		0 & 0 & 0 & 0 \\ 
		-{\rm i}C_{0}V_{0}{\rm e}^{{\rm i}(\theta +\phi )} & 0 & V_{0}^{2} & V_{0}\sqrt{1-V_{0}^{2}}%
		{\rm e}^{{\rm i}\phi } \\ 
		-{\rm i}C_{0}\sqrt{1-V_{0}^{2}}{\rm e}^{{\rm i}\theta } & 0 & V_{0}\sqrt{1-V_{0}^{2}}{\rm e}^{-{\rm i}\phi
		} & 1-V_{0}^{2}%
		\end{array}%
		\right) .\text{ }
		\end{equation}%
	The entanglement between the interfering qubit and the WPD is determined by
	the eigenvalues of the operator $\overset{\sim }{\rho }=\rho \left( \sigma
	_{{\rm q},y}\otimes \sigma _{{\rm w},y}\right) \rho ^{\ast }\left( \sigma _{{\rm q},y}\otimes
	\sigma _{{\rm w},y}\right) $, where $\sigma _{{\rm q},y}$ and $\sigma _{{\rm w},y}$ are the
	y-components of the Pauli operator for the qubit and the WPD, respectively.
	The matrix $\overset{\sim }{\rho }$ reads as 
		\begin{equation}
		\overset{\sim }{\rho }=\frac{1}{4}\left( 
		\begin{array}{cccc}
		\left( 1+C_{0}^{2}\right) \left( 1-V_{0}^{2}\right) & -V_{0}\sqrt{1-V_{0}^{2}%
		}\left( 1+C_{0}^{2}\right) {\rm e}^{{\rm i}\phi } & 0 & 2{\rm i}C_{0}\sqrt{1-V_{0}^{2}}%
		{\rm e}^{-{\rm i}\theta } \\ 
		0 & 0 & 0 & 0 \\ 
		-2{\rm i}V_{0}\left( 1-V_{0}^{2}\right) C_{0}{\rm e}^{{\rm i}(\phi +\theta )} & 2{\rm i}V_{0}^{2}%
		\sqrt{1-V_{0}^{2}}C_{0}{\rm e}^{{\rm i}(2\phi +\theta )} & 0 & V_{0}\sqrt{1-V_{0}^{2}}%
		\left( 1+C_{0}^{2}\right) {\rm e}^{{\rm i}\phi } \\ 
		-2{\rm i}\left( 1-V_{0}^{2}\right) ^{3/2}C_{0}{\rm e}^{{\rm i}\theta } & 2{\rm i}V_{0}\sqrt{%
			1-V_{0}^{2}}C_{0}{\rm e}^{{\rm i}(\phi +\theta )} & 0 & \left( 1+C_{0}^{2}\right) \left(
		1-V_{0}^{2}\right)%
		\end{array}%
		\right) .\text{ }
		\end{equation}%
	The four eigenvalues of $\overset{\sim }{\rho }$ in the decreasing order are
	respectively $\lambda _{1}=\frac{1}{4}\left( 1-V_{0}^{2}\right) \left(
	1+C_{0}\right) ^{2}$, $\lambda _{2}=\frac{1}{4}\left( 1-V_{0}^{2}\right)
	\left( 1-C_{0}\right) ^{2}$ and $\lambda _{3}=\lambda _{4}=0$. The
	corresponding concurrence \cite{wootters_prl1998}, defined as $E=\max \{\sqrt{\lambda _{1}}-%
	\sqrt{\lambda _{2}}-\sqrt{\lambda _{3}}-\sqrt{\lambda _{4}}$, $0\}$, is $%
	C_{0}\sqrt{1-V_{0}^{2}}$.
	
	For the unbalanced interferometer where the two paths are not
	equally populated, the corresponding qubit's state before interaction with
	the WPD can be described by the density operator
	    \begin{equation}
	    \rho _{{\rm q},0}=\cos ^{2}\alpha | 0\rangle \langle
	    0| +\sin ^{2}\alpha | 1\rangle \langle
	    1| +\frac{{\rm i}}{2}C_{0}\left( {\rm e}^{-{\rm i}\theta }| 0\rangle
    	\langle 1| -{\rm e}^{{\rm i}\theta }| 1\rangle
	    \langle 0| \right) .
	    \end{equation}%
	After the qubit-WPD coupling, the total density operator is
\begin{eqnarray}
	\rho &=&\cos ^{2}\alpha | 0\rangle \langle 0|
	\otimes | W_{0}\rangle \langle W_{0}| +\sin
	^{2}\alpha | 1\rangle \langle 1| \otimes
	\left( V_{0}{\rm e}^{{\rm i}\phi }| W_{0}\rangle +\sqrt{1-V_{0}^{2}}
	| W_{1}\rangle \right) \left( \langle W_{0}|
	V_{0}{\rm e}^{-{\rm i}\phi }+\langle W_{1}| \sqrt{1-V_{0}^{2}}\right) \nonumber\\
	&&+\frac{{\rm i}}{2}C_{0}{\rm e}^{-{\rm i}\theta }| 0\rangle \langle
	1| \otimes | W_{0}\rangle \left( \langle
	W_{0}| V_{0}{\rm e}^{-{\rm i}\phi }+\langle W_{1}| \sqrt{%
		1-V_{0}^{2}}\right) \nonumber\\
	&& -\frac{{\rm i}}{2}C_{0}{\rm e}^{{\rm i}\theta }| 1\rangle \langle
	0| \otimes \left( V_{0}{\rm e}^{{\rm i}\phi }| W_{0}\rangle +
	\sqrt{1-V_{0}^{2}}| W_{1}\rangle \right)  \langle
	W_{0}|,
\end{eqnarray}
	which can be rewritten as%
		\begin{equation}
		\rho =\left( 
		\begin{array}{cccc}
		\cos ^{2}\alpha & 0 & \frac{{\rm i}}{2}C_{0}V_{0}{\rm e}^{-{\rm i}(\theta +\phi )} & \frac{{\rm i}}{2%
		}C_{0}\sqrt{1-V_{0}^{2}}{\rm e}^{-{\rm i}\theta } \\ 
		0 & 0 & 0 & 0 \\ 
		-\frac{{\rm i}}{2}\sin ^{2}\alpha C_{0}V_{0}{\rm e}^{{\rm i}(\theta +\phi )} & 0 & \sin
		^{2}\alpha V_{0}^{2} & \sin ^{2}\alpha V_{0}\sqrt{1-V_{0}^{2}}{\rm e}^{{\rm i}\phi } \\ 
		-\frac{{\rm i}}{2}\sin ^{2}\alpha C_{0}\sqrt{1-V_{0}^{2}}{\rm e}^{{\rm i}\theta } & 0 & \sin
		^{2}\alpha V_{0}\sqrt{1-V_{0}^{2}}{\rm e}^{-{\rm i}\phi } & \sin ^{2}\alpha \left(
		1-V_{0}^{2}\right)%
		\end{array}%
		\right) .\text{ }
		\end{equation}%
	The corresponding $\overset{\sim }{\rho }$ matrix is given by
		\begin{equation}
		\overset{\sim }{\rho }=\left( 
		\begin{array}{cccc}
		\frac{1}{4}\left( C_{0}^{2}+\sin ^{2}2\alpha \right) \left(
		1-V_{0}^{2}\right)  & -\frac{1}{4}V_{0}\sqrt{1-V_{0}^{2}}\left( \sin
		^{2}2\alpha +C_{0}^{2}\right) {\rm e}^{{\rm i}\phi } & 0 & {\rm i}\cos ^{2}\alpha C_{0}\sqrt{%
			1-V_{0}^{2}}{\rm e}^{-{\rm i}\theta } \\ 
		0 & 0 & 0 & 0 \\ 
		-{\rm i}\sin ^{2}\alpha V_{0}\left( 1-V_{0}^{2}\right) C_{0}{\rm e}^{{\rm i}(\phi +\theta )} & 
		{\rm i}\sin ^{2}\alpha V_{0}^{2}\sqrt{1-V_{0}^{2}}C_{0}{\rm e}^{{\rm i}(2\phi +\theta )} & 0 & 
		\frac{1}{4}V_{0}\sqrt{1-V_{0}^{2}}\left( \sin ^{2}2\alpha +C_{0}^{2}\right)
		{\rm e}^{{\rm i}\phi } \\ 
		-{\rm i}\sin ^{2}\alpha \left( 1-V_{0}^{2}\right) ^{3/2}C_{0}{\rm e}^{{\rm i}\theta } & {\rm i}\sin
		^{2}\alpha V_{0}\sqrt{1-V_{0}^{2}}C_{0}{\rm e}^{{\rm i}(\phi +\theta )} & 0 & \frac{1}{4}%
		\left( \sin ^{2}2\alpha +C_{0}^{2}\right) \left( 1-V_{0}^{2}\right) 
		\end{array}	
			\right). 
		\end{equation}%
	The four eigenvalues of $\overset{\sim }{\rho }$ in the decreasing order are
	respectively $\lambda _{1}=\frac{1}{4}\left( 1-V_{0}^{2}\right) \left( \sin
	2\alpha +C_{0}\right) ^{2}$, $\lambda _{2}=\frac{1}{4}\left(
	1-V_{0}^{2}\right) \left( \sin 2\alpha -C_{0}\right) ^{2}$ and $\lambda
	_{3}=\lambda _{4}=0$. The corresponding concurrence, defined as $E=\max \{%
	\sqrt{\lambda _{1}}-\sqrt{\lambda _{2}}-\sqrt{\lambda _{3}}-\sqrt{\lambda
		_{4}}$, $0\}$, is also equal to $C_{0}\sqrt{1-V_{0}^{2}}$.

	\section{Supplementary Note 2: Fringe contrast}
	
	After the second Hadamard transformation on the interfering qubit, the total
	density operator is%
		\begin{eqnarray}
			\rho &=&\frac{1}{2}\cos ^{2}\alpha \left( |0\rangle
			-{\rm i}| 1\rangle \right) \left( \langle 0|
			+\langle 1| {\rm i}\right) \otimes | W_{0}\rangle
			\langle W_{0}| \nonumber\\
			&&+\frac{1}{2}\sin ^{2}\alpha \left( | 1\rangle -{\rm i}|
			0\rangle \right) \left( \langle 1| +\langle
			0| {\rm i}\right) \otimes \left( V_{0}{\rm e}^{{\rm i}\phi }|
			W_{0}\rangle +\sqrt{1-V_{0}^{2}}| W_{1}\rangle \right)
			\left( \langle W_{0}| V_{0}{\rm e}^{-{\rm i}\phi }+\langle
			W_{1}| \sqrt{1-V_{0}^{2}}\right) \nonumber\\
			&&+\frac{{\rm i}}{4}C_{0}{\rm e}^{-{\rm i}\theta }\left( | 0\rangle
			-{\rm i}| 1\rangle \right) \left( \langle 1|
			+\langle 0| {\rm i}\right) \otimes | W_{0}\rangle
			\left( \langle W_{0}| V_{0}{\rm e}^{-{\rm i}\phi }+\langle
			W_{1}| \sqrt{1-V_{0}^{2}}\right) \nonumber\\
			&& -\frac{{\rm i}}{4}C_{0}{\rm e}^{{\rm i}\theta }\left( | 1\rangle
			-{\rm i}| 0\rangle \right) \left( \langle 0|
			+\langle 1| {\rm i}\right) \otimes \left( V_{0}{\rm e}^{{\rm i}\phi }|
			W_{0}\rangle +\sqrt{1-V_{0}^{2}}| W_{1}\rangle \right)
			 \langle W_{0}|.
		\end{eqnarray}%
	Tracing over the degree of freedom of the WPD, we obtain the reduced density
	operator of the interfering qubit%
	\begin{eqnarray}
		\rho _{{\rm q}} &=&\frac{1}{2}\cos ^{2}\alpha \left( | 0\rangle
		-{\rm i}| 1\rangle \right) \left( \langle 0|
		+\langle 1| {\rm i}\right) \nonumber\\
		&&+\frac{1}{2}\sin ^{2}\alpha \left( | 1\rangle -{\rm i}|
		0\rangle \right) \left( \langle 1| +\langle
		0| {\rm i}\right)  \nonumber\\
		&&+\frac{{\rm i}}{4}C_{0}V_{0}{\rm e}^{-{\rm i}(\theta +\phi )}\left( |
		0\rangle -{\rm i}| 1\rangle \right) \left( \langle
		1| +\langle 0| {\rm i}\right) \nonumber\\
		&& -\frac{{\rm i}}{4}C_{0}V_{0}{\rm e}^{-{\rm i}(\theta +\phi )}\left( |
		1\rangle -{\rm i}| 0\rangle \right) \left( \langle
		0| +\langle 1| {\rm i}\right)  .
	\end{eqnarray}%
	The probability of detecting the qubit in the state $|
	1\rangle $ is 
	\begin{equation}
	P_{1}=\langle 1| \rho _{{\rm q}}| 1\rangle =\frac{1}{%
		2}\left[ 1+C_{0}V_{0}\cos \left( \theta +\phi \right) \right] .
	\end{equation}%
	Therefore, the fringe contrast is also given by $C_{0}V_{0}$.
	
	\section{Supplementary Note 3: Distinguishability}
	
	The amount of the which-path information stored in the WPD, is quantified by
	the distinguishability, defined as \cite{englert_prl1996} 
	\begin{equation}
     D=\frac{1}{2}\mathrm{tr}_{{\rm w}}| \rho _{{\rm w},0}-\rho _{{\rm w},1}| ,
    \end{equation}%
	where $\rho _{{\rm w},0}$ and $\rho _{{\rm w},1}$ respectively denote the density
	operators of the WPD associated with the interfering qubit's $|
	0\rangle $ and $| 1\rangle $ states after their coupling, given by 
		\begin{eqnarray}
			\rho _{{\rm w},0} &=&\langle 0| \rho | 0\rangle /%
			\mathrm{tr}_{{\rm w}}\left( \langle 0| \rho |
			0\rangle \right) =| W_{0}\rangle \langle
			W_{0}| , \\
			\rho _{{\rm w},1} &=&\langle 1| \rho | 1\rangle /%
			\mathrm{tr}_{{\rm w}}\left( \langle 0| \rho |
			0\rangle \right) =\left( V_{0}{\rm e}^{{\rm i}\phi }| W_{0}\rangle +%
			\sqrt{1-V_{0}^{2}}| W_{1}\rangle \right) \nonumber\\
			&& \left( \langle W_{0}| V_{0}{\rm e}^{-{\rm i}\phi }+\langle W_{1}| \sqrt{%
				1-V_{0}^{2}}\right) .
		\end{eqnarray}%
	In the basis \{$| W_{0}\rangle $, $|
	W_{1}\rangle $\}, $\rho _{{\rm w},0}-\rho _{{\rm w},1}$ is given by the matrix 
	\begin{equation}
	\left( 
	\begin{array}{cc}
	1-V_{0}^{2} & -V_{0}\sqrt{1-V_{0}^{2}}{\rm e}^{{\rm i}\phi } \\ 
	-V_{0}\sqrt{1-V_{0}^{2}}{\rm e}^{-{\rm i}\phi } & V_{0}^{2}-1%
	\end{array}%
	\right) .
	\end{equation}%
	The two eigenvalues of $\rho _{{\rm w},0}-\rho _{{\rm w},1}$ are 
	\begin{equation}
	\lambda _{\pm }=\pm \sqrt{1-V_{0}^{2}}.
	\end{equation}%
	Therefore, the distinguishability is 
	\begin{equation}
	D=\frac{1}{2}\left( | \lambda _{+}| +| \lambda
	_{-}| \right) =\sqrt{1-V_{0}^{2}},
	\end{equation}%
	which is equal to the qubit-WPD concurrence only when $C_{0}=1$.

\section{Supplementary Note 4: Implementation of the dynamic model with additional phases -- Single qubit rotation}
	
	A rotation around the axis 
	\begin{align}
	\textbf{\textit{n}} =\left( \sin \tilde{\theta } \cos \phi ,\sin \tilde{\theta } \sin \phi ,\cos \tilde{\theta } \right), 
	\end{align}
	in which $\tilde{\theta } $ is the polar angle and $\phi$ is the equatorial angle, by an angle $\gamma$,  can be written as a rotation operator 

	\begin{align}
		R^{\textbf{\textit{n}} }_{\gamma} = \begin{pmatrix} \cos \frac{\gamma }{2} -{\rm i}\sin \frac{\gamma }{2} \cos \tilde{\theta } & -{\rm i}\sin \frac{\gamma }{2} \sin \tilde{\theta } {\rm e}^{-{\rm i}\phi } \\ 
			-{\rm i}\sin \frac{\gamma }{2} \sin \tilde{\theta } {\rm e}^{{\rm i}\phi } & \cos \frac{\gamma }{2} +{\rm i}\sin \frac{\gamma }{2} \cos \tilde{\theta } 		\end{pmatrix}.
	\end{align}

	We choose 
	\begin{align}
	&\tilde{\theta } = \frac{\pi }{2} \left(\tilde{\theta } =  \frac{\pi }{2} \right) , \\
	&\phi =  \theta \left( \phi = -\omega_{\rm r} \frac{\pi }{2g_{2}} \right) , \\
	&\gamma = \frac{\pi }{2} \left( \gamma = \frac{\pi }{2} \right),  
	\end{align}
	and apply this rotation operator to qubit $Q_1 \left( Q_{2}\right) $. Thus we have
	\begin{align}
	R^{\textbf{\textit{n}}_{1} }_{\frac{\pi }{2} }|0_1\rangle &=\frac{1}{\sqrt{2} } \begin{pmatrix}1-{\rm i}\cos \tilde{\theta } &-{\rm i}\sin \tilde{\theta } {\rm e}^{-{\rm i}\phi }\\ -{\rm i}\sin \tilde{\theta } {\rm e}^{{\rm i}\phi }&1+\cos \tilde{\theta } \end{pmatrix} \begin{pmatrix}1\\ 0\end{pmatrix} \notag \\
	&\equiv \frac{1}{\sqrt{2} } \left[ |0_1\rangle -{\rm i}{\rm e}^{{\rm i}\theta }|1_1\rangle \right],  \\
	R^{\textbf{\textit{n}}_{2} }_{\frac{\pi }{2} }|0_2\rangle &=\frac{1}{\sqrt{2} } \left[ |0_2\rangle -{\rm i}{\rm e}^{-{\rm i}\omega_{\rm r} \frac{\pi }{2g_{2}} }|1_2\rangle \right].
	\end{align}
	The phase factor $ \phi = -\omega_{\rm r} \frac{\pi }{2g_{2}} $ is chosen to cancel the relative phase induced during the resonant $Q_2$-$R$ interaction. The sum of the modulus of the off-diagonal elements of
	\begin{align} 
	 R^{\textbf{\textit{n}}_{1} }_{\frac{\pi }{2} }|0_1\rangle \langle 0_1|R^{\textbf{\textit{n}}_{1} \dagger}_{\frac{\pi }{2} }=\frac{1}{2} 
	   \begin{pmatrix}1
	     &{\rm i}{\rm e}^{-{\rm i}\theta }\\ -{\rm i}{\rm e}^{{\rm i}\theta }&1
	   \end{pmatrix} 
	 \end{align} 
	 characterizes the coherence, that is 
	\begin{align}
	C_{0}=|\frac{1}{2} {\rm i}{\rm e}^{-{\rm i}\theta }|+|-\frac{1}{2}{\rm i}{\rm e}^{{\rm i}\theta }| =1, 
	\end{align}
	between the two paths. 
	
\section{Supplementary Note 5: Implementation of the dynamic model with additional phases -- $Q_1$-resonator interaction}
	
	Once the $Q_2$-$R$ interaction stops, we tune $Q_2$ back to its idle frequency and adjust $Q_1$ to its working frequency to start the $Q_1$-$R$ interaction, for which only the qubit's transition $|1\rangle \leftrightarrow |2\rangle$ is involved while the other transitions are decoupled. The ideal model describing the interaction is written as:
	\begin{equation}\label{es36}
	H_{Q_{1}R}/\hbar=\omega_{1} |1\rangle \langle 1|+\omega_{2} |2\rangle \langle 2|+\omega_{\rm r} a^{\dagger}a+\sqrt{2}g_{1}\left( a^{\dagger}|1\rangle \langle 2|+a|2\rangle \langle 1|\right),  
	\end{equation}
	where $\hbar\omega_{2} \left( \hbar\omega_{1} \right)  $ is the energy level for the $Q_1$'s $|2\rangle \left( |1\rangle  \right)$ state.
	In the single-excitation subspace $\{|2\rangle|0_{\rm r}\rangle, |1\rangle|1_{\rm r}\rangle\}$, Eq. (\ref{es36}) possesses two eigenstates
	\begin{equation}\label{es37}
	|\phi_{\pm}\rangle = N_{\pm}^{-1}\left[2\sqrt{2}g_1|1\rangle|1_{\rm r}\rangle+(\delta\pm 2\Omega)|2\rangle|0_{\rm r}\rangle\right],
	\end{equation}
	where $N_{\pm}=\sqrt{(\delta\pm2\Omega)^2+8g_1^2}$ are the normalization factors, $\delta=\omega_2-\omega_1-\omega_{{\rm r}}$ and $\Omega=\sqrt{2g_1^2+\delta^2/4}$. The corresponding eigenenergies for Eq. (\ref{es37}) are
	\begin{equation}
	E_{\pm}=\hbar(2\omega_{2}-\delta)/2 \pm \hbar\Omega.
	\end{equation} 
	
	The temporal evolution of the initial qubit-resonator state $|\Psi'(0)\rangle=|1\rangle|1_{\rm r}\rangle$ can be expressed as
	\begin{eqnarray}
	|\Psi(t)\rangle&=&{\rm e}^{-{\rm i}H_{Q_1R}t/\hbar}|\Psi(0)\rangle \equiv {\rm e}^{-{\rm i}H_{Q_1R}t/\hbar} \frac{1}{8\sqrt{2}g_1\Omega} \left[N_{-}(\delta+2\Omega)	|\phi_{-}\rangle-N_{+}(\delta-2\Omega)	|\phi_{+}\rangle \right] \nonumber \\
	&=&  \frac{1}{8\sqrt{2}g_1\Omega} \left[N_{-}(\delta+2\Omega){\rm e}^{-{\rm i}E_{-}t}	|\phi_{-}\rangle - N_{+}(\delta-2\Omega){\rm e}^{-{\rm i}E_{+}t}	|\phi_{+}\rangle \right] \nonumber \\
	&=& {\rm e}^{-{\rm i}\frac{2\omega_{2}-\delta}{2}t}\left\{ \left[ \cos(\Omega t)+{\rm i}\frac{\delta}{2\Omega} \sin(\Omega t)\right]|1\rangle|1_{\rm r}\rangle-{\rm i}\frac{\sqrt{2}g_1}{\Omega}\sin(\Omega t)|2\rangle|0_{\rm r}\rangle\right\},
	\end{eqnarray}
	which at $t=\pi/\Omega$ reaches 
	\begin{equation}
	|\Psi(\pi/\Omega)\rangle = {\rm e}^{{\rm i}\pi\left(1-\frac{2\omega_{2}-\delta}{2\Omega}\right)}|1\rangle|1_{\rm r}\rangle. 
	\end{equation}
	In the other words, the resonator undergoes an evolution 
	conditional on $Q_1$' state $|1\rangle$, i.e.,
	\begin{equation}
	U={\rm e}^{{\rm i}\beta |1_{\rm r}\rangle\langle 1_{{\rm r}}|},
	\end{equation}
	with the corresponding phase factor
	\begin{equation}
	\beta=\pi\left(1-\frac{2\omega_{2}-\delta}{2\Omega}\right).
	\end{equation}  
	In the rotating frame $H = (\omega_{1}+\omega_{\rm r})( |1\rangle  |1_{\rm r}\rangle+|2\rangle |0_{\rm r}\rangle) $, the conditional phase reads 
	\begin{equation}
	\beta=\pi\left[1-\delta/(2\Omega)\right],
	\end{equation} 
	as presented in the main text.
			
	Now, if we start from the initial qubit-resonator state:
	\begin{equation}
	|\Psi^{\prime} \left( 0\right)  \rangle =\frac{1}{\sqrt{2} } \left( |0\rangle -{\rm i}{\rm e}^{{\rm i}\theta }|1\rangle \right)  |W_{0}\rangle,
	\end{equation}
	after the qubit-resonator coupling for an interaction time $t=\frac{\pi }{\Omega } $, we get
	\begin{eqnarray}
	|\Psi^{\prime} \left( \pi/\Omega \right)  \rangle =\frac{1}{\sqrt{2} } \left[ |0\rangle \frac{1}{\sqrt{2} } \left( |0_{\rm r}\rangle -{\rm e}^{-{\rm i}\omega_{\rm r} \frac{\pi }{\Omega } }|1_{\rm r}\rangle \right)  -{\rm i}{\rm e}^{{\rm i}\theta }|1\rangle \frac{1}{\sqrt{2} } \left( {\rm e}^{-{\rm i}\omega_{1} \frac{\pi }{\Omega } }|0_{\rm r}\rangle -{\rm e}^{{\rm i}\beta }|1\rangle_{\rm r} \right)  \right],
	\end{eqnarray}
	which can be rewritten as
	\begin{eqnarray}\label{es38}
	|\Psi^{\prime} \left(\pi/\Omega  \right)  \rangle &=&\frac{1}{\sqrt{2} } \left[ |0\rangle \frac{1}{\sqrt{2} } \left( |0_{\rm r}\rangle -{\rm e}^{-{\rm i}\omega_{\rm r} \frac{\pi }{\Omega } }|1_{\rm r}\rangle \right)  -{\rm i}{\rm e}^{{\rm i}\theta }{\rm e}^{-{\rm i}\omega_{1} \frac{\pi }{\Omega } }|1\rangle \frac{1}{\sqrt{2} } \left( |0_{\rm r}\rangle -{\rm e}^{{\rm i}\left( \beta +\omega_{1} \frac{\pi }{\Omega } \right)  }|1_{\rm r}\rangle \right)  \right] \notag \\  
	&=&\frac{1}{\sqrt{2} } \left[ |0\rangle |W^{\prime }_{0}\rangle -{\rm i}{\rm e}^{{\rm i}\left( \theta -\omega_{1} \frac{\pi }{\Omega } \right)  }|1\rangle U^{\prime }|W^{\prime }_{0}\rangle \right],
	\end{eqnarray}
	where we define
	\begin{align}
	|W^{\prime }_{0}\rangle&=\frac{1}{\sqrt{2} } \left( |0_{\rm r}\rangle -{\rm e}^{-{\rm i}\omega_{\rm r} \frac{\pi }{\Omega } }|1_{\rm r}\rangle \right), \\
	U^{\prime }&={\rm e}^{{\rm i}\beta^{\prime } |1_{\rm r}\rangle \langle 1_{{\rm r}}|}, \\
	\beta^{\prime } &=\beta +\omega_{1} \frac{\pi }{\Omega } +\omega_{\rm r} \frac{\pi }{\Omega }. 
	\end{align}

	\section{Supplementary Note 6: Implementation of the dynamic model with additional phases -- The second Hadamard operation}
	
	The dynamics for the Hadamard operation is modeled by the interaction Hamiltonian 
	\begin{equation}\label{es31}
	H_q = \frac{1}{2}\hbar \varepsilon {\rm e}^{{\rm i}\varphi}|1\rangle \langle 0| + h.c,                                                                                                                        
	\end{equation}
	where $\varepsilon$, $\varphi$ and $t$ denote the amplitude, phase and delay of the driving pulse, respectively. Eq. (\ref{es31}) leads to the following state evolutions
	\begin{equation}
	|0\rangle \rightarrow \cos(\frac{\varepsilon}{2}t)|0\rangle - {\rm i}{\rm e}^{{\rm i}\varphi}\sin(\frac{\varepsilon}{2}t)|1\rangle,
	\end{equation} 
	\begin{equation}
	|1\rangle \rightarrow \cos(\frac{\varepsilon}{2}t)|1\rangle - {\rm i}{\rm e}^{-{\rm i}\varphi}\sin(\frac{\varepsilon}{2}t)|0\rangle.
    \end{equation}
	The Hadamard operation is given by $R^{\textbf{\textit{x}} }_{\frac{\pi }{2} }$, which is a single qubit rotation around the $x$-axis by an angle $\pi/2$, corresponding to the transformation ($\varphi=0$, $\varepsilon t = \pi/2$)
	\begin{equation}
	|0\rangle \rightarrow \left( |0\rangle -{\rm i}|1\rangle \right)/{\sqrt{2}},
	\end{equation}
    \begin{equation}
	|1\rangle \rightarrow \left( |1\rangle -{\rm i}|0\rangle \right)/\sqrt{2}.
	\end{equation}
	After the second Hadamard operation, the state, i.e., Eq. (\ref{es38}), of the $Q_1$-$R$ coupled system becomes 
	\begin{eqnarray}\label{es39}
	R^{\textbf{\textit{x}} }_{\frac{\pi }{2} }|\Psi^\prime \left( \pi \text{/} \Omega \right) \rangle &=&\frac{1}{2} \left[ \left( |0\rangle -{\rm i}|1\rangle \right)  |W^{\prime }_{0}\rangle -{\rm i}{\rm e}^{{\rm i}\left( \theta -\omega_{1} \frac{\pi }{\Omega } \right)  }\left( |1\rangle -{\rm i}|0\rangle \right)  U^{\prime }|W^{\prime }_{0}\rangle \right] \notag \\
	&=&\frac{1}{2} \left[ |0\rangle \left( |W^{\prime }_{0}\rangle -{\rm e}^{{\rm i}\left( \theta -\omega_{1} \frac{\pi }{\Omega } \right)  }U^{\prime }|W^{\prime }_{0}\rangle \right) -{\rm i}|1\rangle \left( |W^{\prime }_{0}\rangle +{\rm e}^{{\rm i}\left( \theta -\omega_{1} \frac{\pi }{\Omega } \right)  }U^{\prime }|W^{\prime }_{0}\rangle \right)  \right]  \label{get P1 from}.
	\end{eqnarray}
	With Eq. (\ref{es39}), we can deduce that the probability of detecting $Q_1$ in the state $|1\rangle$ is
	\begin{eqnarray}
	P_{1}^{\prime}&=&\frac{1}{4} \left( \langle W^{\prime }_{0}|+{\rm e}^{-{\rm i}\left( \theta -\omega_{1} \frac{\pi }{\Omega } \right)  }\langle W^{\prime }_{0}|U^{\prime^{\dagger} }\right)  \left( |W^{\prime }_{0}\rangle +{\rm e}^{{\rm i}\left( \theta -\omega_{1} \frac{\pi }{\Omega } \right)  }U^{\prime }|W^{\prime }_{0}\rangle \right) \notag \\
	&=&\frac{1}{4} \left( 1+{\rm e}^{-{\rm i}\left( \theta -\omega_{1} \frac{\pi }{\Omega } \right)  }\langle W^{\prime }_{0}|U^{\prime^{\dagger} }|W^{\prime }_{0}\rangle +{\rm e}^{{\rm i}\left( \theta -\omega_{1} \frac{\pi }{\Omega } \right)  }\langle W^{\prime }_{0}|U^{\prime }|W^{\prime }_{0}\rangle +1\right) \notag \\
	&=&\frac{1}{2} \left[ 1+V^{\prime }_{0}\cos \left( \theta +\phi^{\prime } -\omega_{1} \frac{\pi }{\Omega } \right)  \right],  
	\end{eqnarray}
	where $V^{\prime }_{0}=\left| \langle W^{\prime }_{0}|U^{\prime }|W^{\prime }_{0}\rangle \right|  $ and $\phi^{\prime } =arg(\langle W^{\prime }_{0}|U^{\prime }|W^{\prime }_{0}\rangle )$.

\section{Supplementary Note 7: Device parameters and experimental setup}
	Our device consists of five frequency-tunable superconducting Xmon qubits, with the anharmonicities being about $\alpha_j \simeq 2\pi\times240$ MHz. Each qubit has a microwave line (XY line) for driving its state transition, and an individual flux line (Z line) for dynamically tuning its frequency \cite{Song_natcomm2017, Barends_nature2014}, allowing its flexible coupling (with the strength $g_j$) to the bus resonator, whose 
	bare frequency and energy relaxation time are $\omega_{\rm r}/2\pi \simeq \SI{5.582} {\giga\hertz}$ and $T_{{\rm r}} \simeq 13$ $\mu$s (see Supplementary Table \ref{table1}), respectively \cite{Song_natcomm2017, Ning_prl2019, Zhenbiao_npj_quantuminf2021, Xu_optica2021}. 
	Besides, each qubit is dispersively coupled to its own readout resonator, whose frequency and leakage rate are $\omega_{{\rm re},j}/2\pi$ and $\kappa_{{\rm r},j}$, respectively. All the readout resonators couple to a common transmission line for multiplexed readout of all qubits' states. The readout is the type of single-shot measurement and is achieved with assistance of an impedance-transformed Josephson parametric amplifier (JPA) with bandwidth of about $150$ MHz used to strength the signal-to-noise ratio. The two qubits chosen in the experiment, are marked as $Q_{1}$ and $Q_{2}$, whose properties are characterized and listed in Supplementary Table \ref{table1}. Qubits' s energy relaxation time $T_{1,j}^{|k\rangle}$ ($j$ and $k$ indicate the qubit and its state, respectively), the Ramsey Gaussian dephasing time $T_{2,j}^{*,|k\rangle}$, and the spin echo Gaussian dephasing time $T_{2,j}^{SE,|k\rangle}$ are measured at their idle frequency $\omega_{{\rm id}}^{|1\rangle}/2\pi$. The probability of correctly reading out each qubit ($Q_j$) in state $\vert k\rangle$ is $F_{k,j}$. 
	
		\begin{figure}[htbp] 
		\centering
		\includegraphics[width=5.0in]{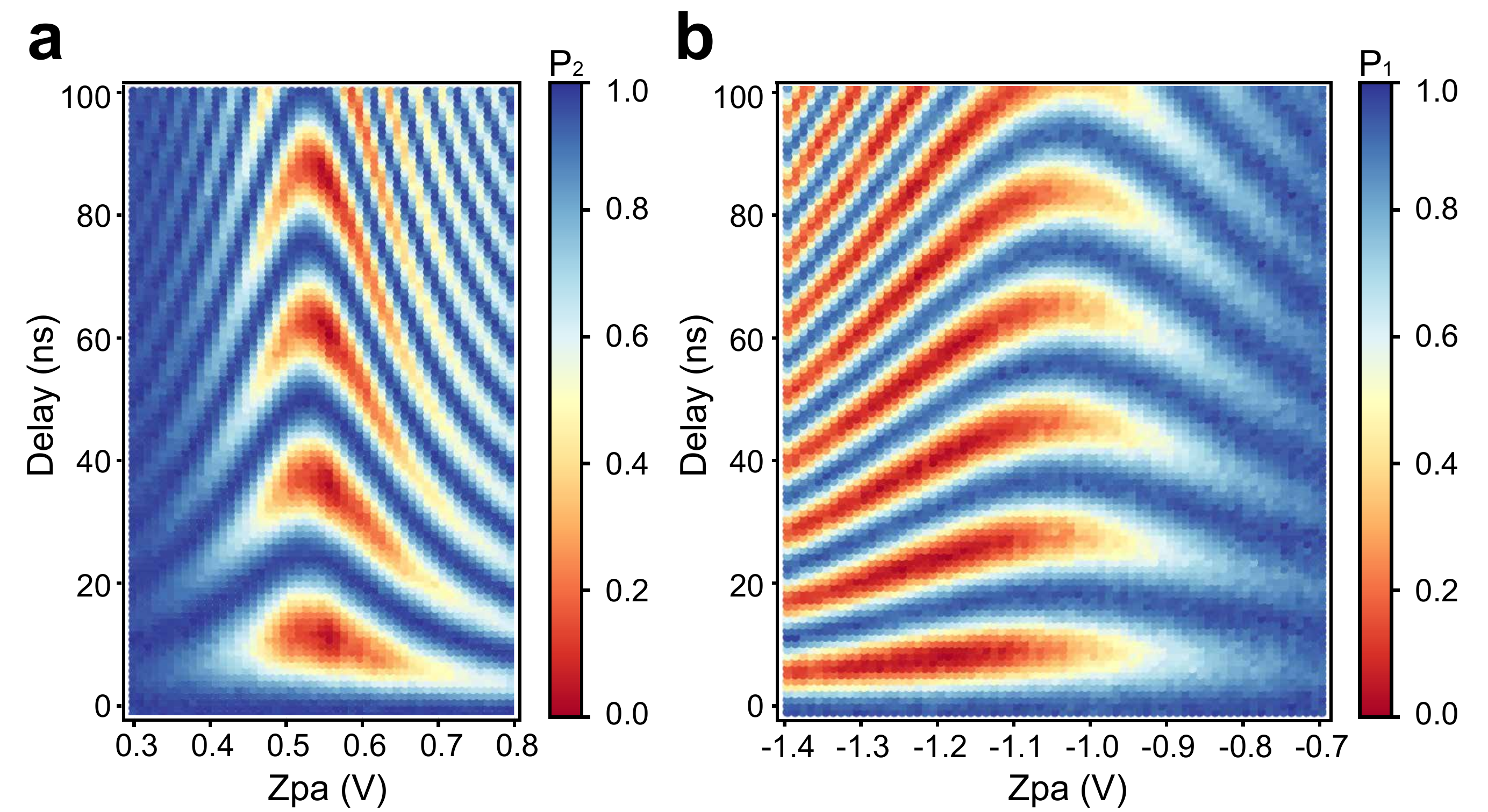}
		\caption{\textbf{Vacuum Rabi swap spectroscopy.} 
			\textbf{a} The swapping between $Q_{2}$'s $|1\rangle |0_{\rm r}\rangle \leftrightarrow |0\rangle |1_{\rm r}\rangle$ transition and the resonator.
			\textbf{b} The swapping between $Q_1$'s $|2\rangle |0_{\rm r}\rangle \leftrightarrow |1\rangle |1_{\rm r}\rangle$ transition and the resonator.}
		\label{QRswap}
	\end{figure}
	
	\begin{table*}
		\centering
		\renewcommand{\arraystretch}{1}
		\begin{tabular}{{ccc}}
			\hline\hline
			\centering
			Parameters & \qquad\qquad\qquad\qquad $Q_{1}$ & \qquad\qquad\qquad\qquad  $Q_{2}$  \qquad\qquad\qquad\qquad \qquad   \\ \hline
			\centering
			Qubit's idle frequency, $\omega_{{\rm id}}^{|1\rangle}/2\pi$ &\qquad\qquad\qquad\qquad  $\SI{5.967}{\giga\hertz}$    &  $\SI{5.354}{\giga\hertz}$ \\
			
			Readout frequency, $\omega_{{\rm re},j}/2\pi$ &\qquad\qquad\qquad\qquad  $\SI{6.850}{\giga\hertz}$    &  $\SI{6.764}{\giga\hertz}$ \\
			
			Coupling strength to the bus resonator, $g_{j}/2\pi$&\qquad\qquad\qquad\qquad  $\SI{19.2}{\mega\hertz} $ ($|0\rangle \leftrightarrow |1\rangle$)   &  $\SI{19.9}{\mega\hertz}$ ($|0\rangle \leftrightarrow|1\rangle$)     \\
			
			Energy relaxation time, $T_{1,j}^{|1\rangle}$ ($T_{1,j}^{|2\rangle}$) &\qquad\qquad\qquad\qquad  $17.1$ $(7.6)$ $\SI{}{\micro\second}$   &$23.4$ $(14.7)$ $\SI{}{\micro\second}$   \\
			
			Ramsey dephasing time, $T^{\star,|1\rangle}_{2,j}$ ($T^{\star,|2\rangle}_{2,j}$)  &\qquad\qquad\qquad\qquad  $3.0$ $(2.3)$ $\SI{}{\micro\second}$   &$2.4$ $(2.2)$ $\SI{}{\micro\second}$     \\
			
			Dephasing time with spin echo, $T^{SE,|1\rangle}_{2,j}$ ($T^{SE,|2\rangle}_{2,j}$) &\qquad\qquad\qquad\qquad  $5.7$ $(4.7)$ $\SI{}{\micro\second}$   &$6.5$ $(3.2)$ $\SI{}{\micro\second}$ \\
			
			Leakage rate of readout resonator, $\kappa_{{\rm r},j}$ &\qquad\qquad\qquad\qquad  $1/(\SI{313}{\nano\second})$   &$1/(\SI{315}{\nano\second})$ \\
			
			Nonlinearity, $\alpha_{j}/2\pi$ &\qquad\qquad\qquad\qquad  $\SI{241}{\mega\hertz}$   &$\SI{240}{\mega\hertz}$ \\
			
			$|0 \rangle$ state readout fidelity, $F_{0,j}$ &\qquad\qquad\qquad\qquad  $ 0.9930$   &$0.9803$  \\
			$|1 \rangle$ state readout fidelity, $F_{1,j}$&\qquad\qquad\qquad\qquad  $0.8917$   &$0.9073$ \\
			$|2 \rangle$ state readout fidelity, $F_{2,j}$&\qquad\qquad\qquad\qquad  $0.8483$   &$0.8890$   \\ \hline\hline
			
		\end{tabular}
			\caption{\label{table1} \textbf{Qubits characteristics.} In the experiment, both qubits are initialized to the ground state $|0\rangle$ at their respective idle frequencies $\omega_{{\rm id},j}/2\pi$, at which point the listed qubit' coherence parameters (including the lifetime $T_{1,j}^{|k\rangle}$, the Ramsey Gaussian dephasing time $T_{2,j}^{*,|k\rangle}$, and the spin echo Gaussian dephasing time $T_{2,j}^{SE,|k\rangle}$) are measured. $\omega_{{\rm id},j}$ is also the point at which single-qubit rotations and state tomographies are performed. $g_1$ ($g_2$) is the coupling strength between $Q_1$ ($Q_2$) and the bus resonator, achieved by measurement of Rabi oscillation between $Q_1$'s $|0\rangle \leftrightarrow |1\rangle$ ($Q_2$'s $|0\rangle \leftrightarrow |1\rangle$) transition and the bus resonator's photon. The detuning between the qubit's $|1\rangle \leftrightarrow |2\rangle$ and $|0\rangle \leftrightarrow |1\rangle$ transitions gives the qubit's anharmonicity $\alpha_j$. The fidelity for correctly measuring each qubit's state is $F_{j,k}$, characterized by extracting information of each readout resonator, whose frequency and leakage rate are $\omega_{{\rm re},j}/2\pi$ and $\kappa_{{\rm r}}$, respectively.} 
		
	\end{table*}
    
    When the bias pulse tunes the qubit frequency exactly on resonance with the resonator, through either $|1\rangle \leftrightarrow |2\rangle$ or $|0\rangle \leftrightarrow |1\rangle$ transition, 
    the vacuum Rabi swaps between the qubit and the resonator have the largest amplitude and the longest duration. The coupling strength can be characterized by fitting the oscillation of $P_{2}$ ($P_{1}$) versus the evolution time at a fixed frequency. Supplementary Figure \ref{QRswap} shows the swap spectroscopy for $Q_{1}$ and $Q_{2}$ by sweeping the bias amplitude and the evolution time. Another spectroscopy measurement is implemented to characterize the qubit frequency as a function of the bias, as shown in Supplementary Figure \ref{Spectroscopy}. The avoided crossing at the resonant point indicates that the 
    coupling strength between $Q_1$'s $\vert 1 \rangle \leftrightarrow \vert 2\rangle$ transition and the resonator is $2\pi \times 27.1$ MHz. 
    The experimental setup where the whole electronics and wiring for the device control is outlined and shown in Supplementary Figure \ref{Schematic lay out of the experimental setup} \cite{Song_natcomm2017, Ning_prl2019, Zhenbiao_npj_quantuminf2021, Xu_optica2021}.       
    
	\begin{figure*}[htbp] 
		\centering
		\includegraphics[width=6.0in]{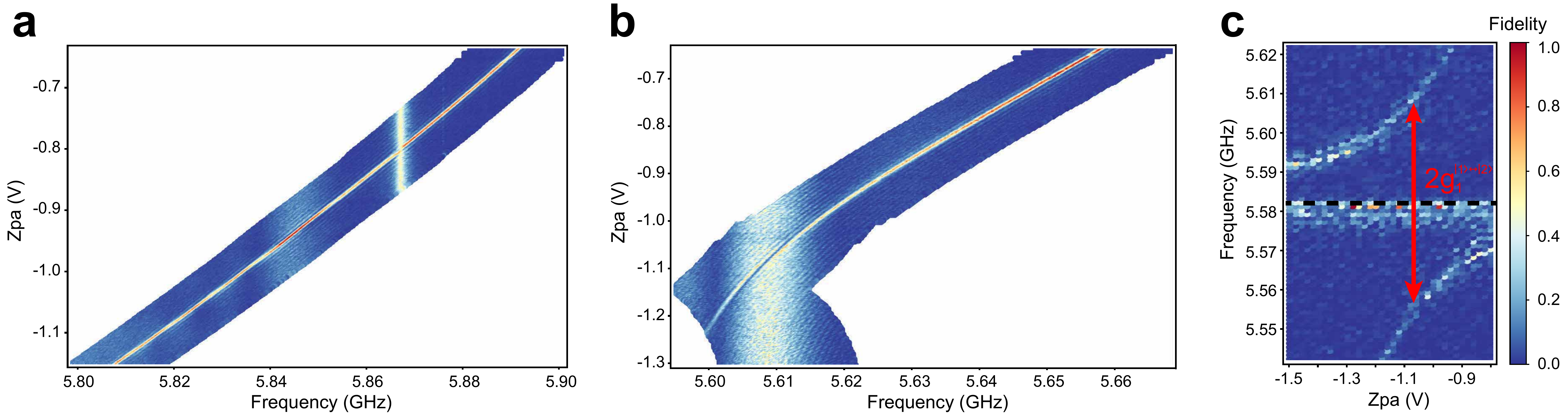}
		\caption{\textbf{The qubit spectroscopy.} 
								\textbf{a} Spectroscopy for $Q_{1}$'s $f_{10}$.
								\textbf{b} Spectroscopy for $Q_{1}$'s $f_{21}$.
								\textbf{c} Spectroscopy for $Q_{1}$'s $f_{21}$. Note that, in the spectroscopy experiment shown in \textbf{c}, we insert a rectangular pulse which swaps the excitation between $Q_{1}$ and the bus resonator before the readout pulse. The black dash line indicates the frequency of the bus resonator. The avoided crossing whose separation reveals the coupling strength between the qubit
								 and the resonator is observed.}
		\label{Spectroscopy}
	\end{figure*}

	\begin{figure*}[htbp] 
		\centering
		\includegraphics[width=5in]{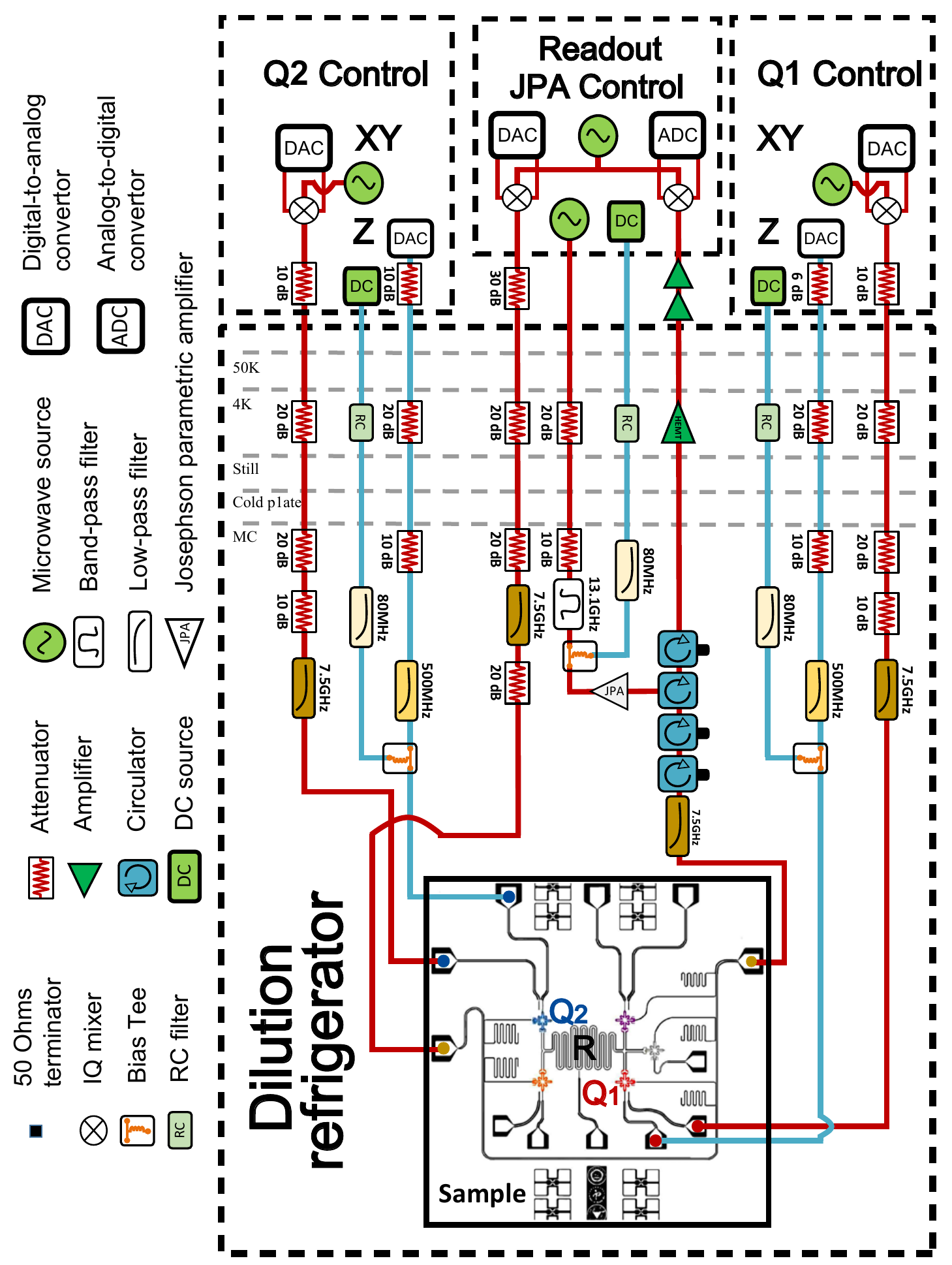}
		\caption{\textbf{Schematic lay out of the device and the experimental setup.} The used superconducting circuit sample possesses five frequency-tunable Xmon qubits, two of which are used and labled as $Q_1$ and $Q_2$; all the qubits are coupled to a bus resonator R, which has a fixed bare frequency $\omega_{{\rm r}}/2\pi =5.582$ GHz; each qubit can be individually frequency-biased and frequency-modulated (through Z line) and flipped (through XY line), and has its own readout resonator assisting to produce its state information. Each qubit's XY control is implemented through the mixing of the low-frequency signals yielded by 2 Digital-to-analog converter (DAC)'s I/Q channels and a Microwave source (MS) whose carrier frequency is about $5.5$ GHz; while the Z control is implemented by two signals: one is produced by the Direct-current (DC) biasing line from a low frequency DC source, the other is directly from the Z control of a DAC. The qubits' readout is implemented by the mixing of the signals of 2 DAC's I/Q channels and a MS with the frequency of about $6.6$ GHz as to output a readout pulse with 2 tones targeting 2 qubits' readout resonators. The output from the sample, before being captured and demodulated by Analog-to-digital converter (ADC), is amplified sequentially by the impedance-transformed Josephson parametric amplifier (JPA, which is pumped by a MS with the frequency of about $13.5$ GHz and modulated by a DC bias), high electron mobility transistor (HEMT) and room temperature amplifiers. Both the DAC and ADC are field programmable gate array (FPGA)-controlled to respond at the nanosecond scale. Furthermore, the custom-made circulators, attenuators and filters are added to the specific locations of the signal lines to reduce the noises affecting the operation of the device.}
		\label{Schematic lay out of the experimental setup}
	\end{figure*}

\section{Supplementary Note 8: Readout correction}
	The fidelity matrix for calibrating the measured probabilities is defined as \cite{Song_natcomm2017,Ning_prl2019,Zhenbiao_npj_quantuminf2021,Xu_optica2021,Chao_PRL2017,Zhengy_PRL2017}
	\begin{equation}
	\hat{F} = \begin{pmatrix}
	f_{0} & e_{01} & e_{02} \\ 
	e_{10} & f_{1} & e_{12} \\
	e_{20} & e_{21} & f_{2} \\
	\end{pmatrix},
	\end{equation}
	where $f_{j}$ ($j=0,1,2$) is the probability that correctly gets the qubit's $|0\rangle$, $|1\rangle$ and $|2\rangle$ state by the measurement, $e_{jk}$ ($j,k=0,1,2$) represents the errors describing the leakage probabilities from state $|k\rangle$ to $|j\rangle$. If we define $\hat{P}_{m}$ as the measured probabilities and $\hat{P}_{i}$ as the intrinsic probabilities for the system. The relation between $\hat{F}$, $\hat{P}_{m}$ and $\hat{P}_{i}$ is simply given by
	\begin{equation}
	\hat{P}_{m} = \hat{F} \cdot \hat{P}_{i}, 
	\end{equation}
	which means the system's intrinsic states are reconstructed by making matrix inversion. The data used in our calibration is extracted from the measured $I$-$Q$ values, as shown in Supplementary Figure \ref{IQraw}.
	
	\begin{figure*}[htbp] 
		\centering
		\includegraphics[width=5.0in]{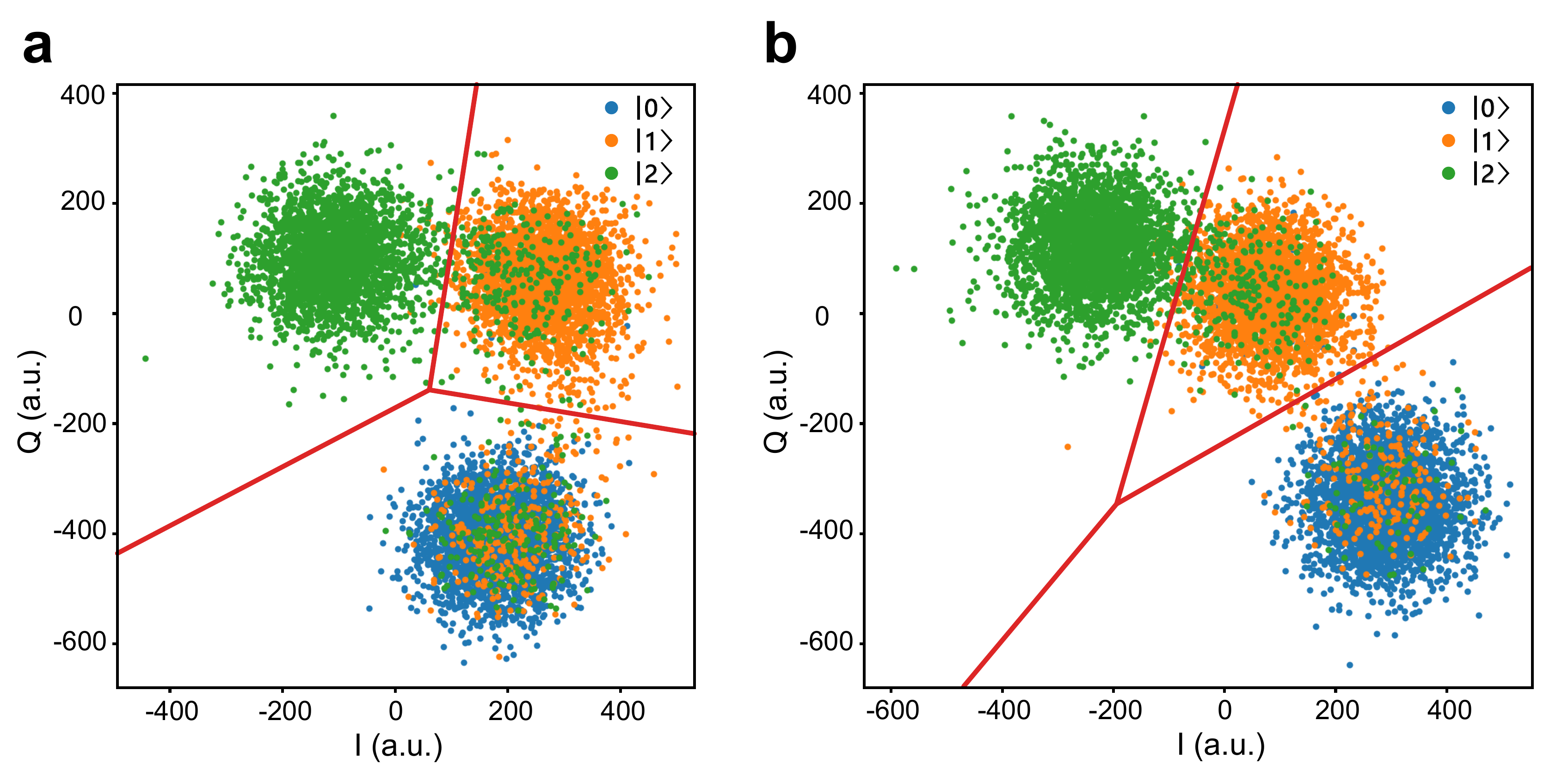}
		\caption{\textbf{The measured $I$-$Q$ values when each qubit is prepared in $\vert 0\rangle$ (blue), $\vert 1\rangle$ (orange) and $\vert 2\rangle$ (green) state.} \textbf{a} The $I$-$Q$ data of $Q_1$. \textbf{b} The $I$-$Q$ data of  qubit Q$_2$.}
		\label{IQraw}
	\end{figure*}

\section{Supplementary Note 9: Numerical Simulation}
	Numerical simulations are performed to verify the experimental results. These simulations mainly consider two Hamiltonians,  
    for $Q_{2}$-$R$ and $Q_{1}$-$R$ interaction, respectively, i.e., 
	\begin{equation}
	H_{Q_2R}/\hbar = \omega_{1,Q_2} |1 \rangle\langle 1|+\omega_{\rm r}a^{\dagger}a +g_{2} ( a^{\dagger}|0\rangle\langle 1| + a|1\rangle\langle 0|),
	\end{equation}
	\begin{eqnarray}
	H_{Q_1R}/\hbar = \omega_{1,Q_1} |1 \rangle\langle 1|+\omega_{2,Q_1}|2 \rangle\langle 2| +\omega_{\rm r}a^{\dagger}a+\sqrt{2}g_{1} ( a^{\dagger}|1\rangle\langle 2| + a|2\rangle\langle 1|),
	\end{eqnarray}
	where $\omega_{j,Q_j}/2\pi$ is the working frequency point for $Q_j$'s $|j\rangle$ state, $a$ and $a^{\dagger}$ are the annihilation and creation operators for photons in the bus resonator, and $|k\rangle\langle l |$ is $Q_j$'s jumping operator from state $|l\rangle$ to $|k\rangle$. Decoherence of $Q_j$ and the bus resonator coupling to environment (consider a Markovian case for simplicity) is included by use of the Lindblad master equation      	
	\begin{equation}
		\dot\rho = -\frac{{\rm i}}{\hbar}[H_{Q_jR},\rho]+\frac{1}{T_{\rm r}} \mathcal D[a]\rho +\frac{1}{T_{1,j}^{|1\rangle}} \mathcal D [b_{j}]\rho +\frac{1}{T_{\phi,j}^{|1\rangle}}  \mathcal D [b_j^{\dagger}b_j]\rho,
	\end{equation}
	where $a$ and $b_j$ are the annihilation operators for the bus resonator and the qubit \cite{Barends_prl2013}, respectively, the qubit dephasing rate is $1/T_{\phi,j}^{|1\rangle} = 1/T_{2,j}^{*,|1\rangle}-1/2T_{1,j}^{|1\rangle}$, the Lindblad super-operator $\mathcal D[O]$ is defined as	
	\begin{equation}
	\mathcal D[O]\rho= O\rho O^{\dagger} -\frac{1}{2}O^{\dagger}O\rho -\frac{1}{2}\rho O^{\dagger}O.
	\end{equation}

\begin{center}
	\bf{Supplementary Note 10: Experimental pulse sequence}
\end{center}
	The pulse sequences for the experiment in the temporal order are shown in Supplementary Figure \ref{Pulse sequence}, which are roughly summed up to include three stages: the preparation of the specific superposition state for the resonator which acts as the WPD; the realization of a conditional phase gate for the resonator's photon state conditional on $Q_1$'s $|1\rangle$ state; the readout of $Q_1$'s state (Supplementary Figure \ref{Pulse sequence}a) and $Q_1$-$Q_2$'s joint states (Supplementary Figure \ref{Pulse sequence}b), which are respectively used to reveal the Ramsey signals and characterize the qubit-WPD entanglement.  

	\begin{figure*}[htbp] 
	\centering
	\includegraphics[width=6.0in]{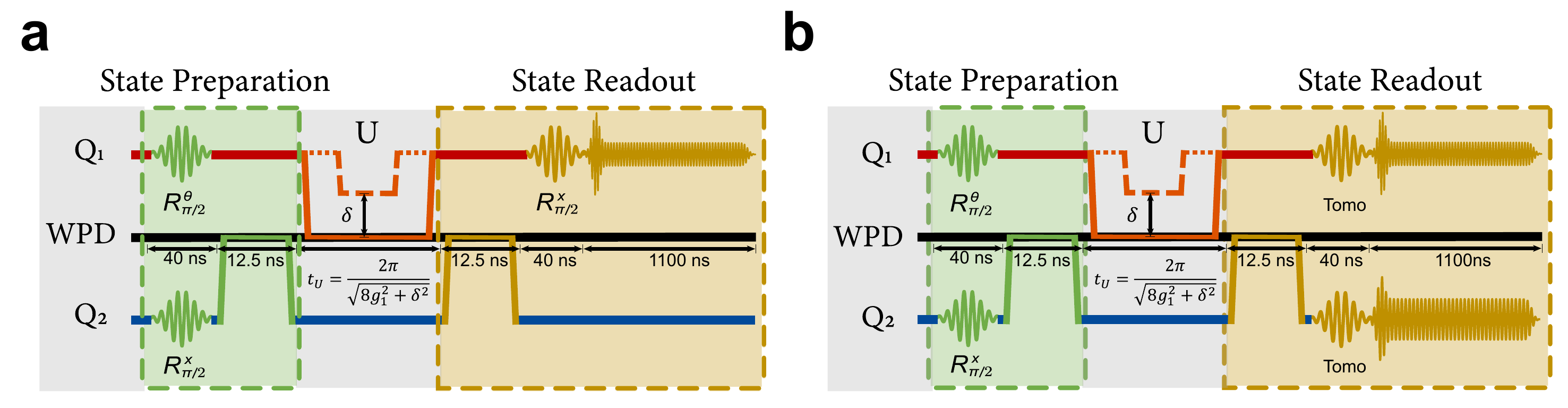}
	\caption{\textbf{Pulse sequence.} 
		\textbf{a} The pulse sequences for detecting the probability of $Q_1$ in $|1\rangle$ state. The creation of the WPD's state: (1) a microwave $\pi/2$-pulse $R_{\pi/2}^x$ prepares $Q_2$ in $(|0\rangle -{\rm i} |1\rangle)/\sqrt{2}$ state; (2) a rectangular pulse then tunes $Q_2$ to the bus resonator (R)'s frequency to couple its $|0\rangle \leftrightarrow |1\rangle$ transition with $R$ for a period of $12.5$ ns before tuning it back to its idle frequency. The realization of a conditional phase gate operation $U$ for the WPD: (1) a microwave $\pi/2$-pulse $R_{\pi/2}^{\theta}$ applied on $Q_1$ prepares it in $(|0\rangle -{\rm i} {\rm e}^{{\rm i}\theta}|1\rangle)/\sqrt{2}$ state; (2) a rectangular pulse then tunes $Q_1$'s $|1\rangle \leftrightarrow |2\rangle$ close to the bus resonator, realizing the $U$-transformation for the resonator's single photon state conditional on $Q_1$'s $|1\rangle$ state; (3) a second microwave $\pi/2$-pulse $R_{\pi/2}^{x}$ applied on $Q_1$ transforming its $|0\rangle$ ($|1\rangle$) state to $(|0\rangle - {\rm i}|1\rangle)/\sqrt{2}$ ($(|1\rangle - {\rm i}|0\rangle)/\sqrt{2}$) state. The final step is the measurement of $Q_1$ to distinguish whether it is in $|1\rangle$ state, as to reveal the Ramsey signals. Before doing this measurement, the bus resonator state is mapped onto $Q_2$: a rectangular pulse tuning $Q_2$'s $|0\rangle \leftrightarrow |1\rangle$ transition to the bus resonator's frequency for a period of $12.5$ ns. \textbf{b} The pulse sequences for measuring the entanglement between $Q_1$ and the resonator. The difference to \textbf{a} is that the joint tomography of both qubits are performed, as to characterize the entanglement between $Q_1$ and the resonator.}  
	\label{Pulse sequence}
\end{figure*}

\newpage
\section{Supplementary References}